\documentclass[aps,pre,twocolumn,superscriptaddress,floatfix,10pt]{revtex4-1}
\usepackage{graphicx}
\usepackage{dcolumn}
\usepackage{bm}
\usepackage{latexsym}
\usepackage{amssymb}
\usepackage{amsmath}
\begin{document}
\title{A Multispecies Exclusion Model Inspired By Transcriptional Interference}
\author{Soumendu Ghosh}
\affiliation{Department of Physics, Indian Institute of Technology
  Kanpur, 208016, India}
\author{Tripti Bameta} \affiliation{UM-DAE Center for Excellence in Basic Sciences, University of Mumbai, Vidhyanagari Campus, Mumbai-400098, India}
\author{Dipanwita Ghanti} 
\affiliation{Department of Physics, Indian
  Institute of Technology Kanpur, 208016, India} 
\author{Debashish Chowdhury}
\email[Corresponding author; E-mail:]{debch@iitk.ac.in}
\affiliation{Department of Physics, Indian Institute of Technology
  Kanpur, 208016, India}
\begin{abstract}
We introduce exclusion models of two distinguishable species of hard rods 
with their distinct sites of entry and exit under open boundary conditions. 
In the first model both species of rods move in the same direction whereas 
in the other two models they move in the opposite direction. These models are 
motivated by the biological phenomenon known as Transcriptional Interference. 
Therefore, the rules for the kinetics of the models, particularly the 
rules for the outcome of the encounter of the rods, are also formulated 
to mimic those observed in Transcriptional Interference. By a combination 
of mean-field theory and computer simulation of these models we demonstrate 
how the flux of one species of rods is completely switched off by the other. 
Exploring the parameter space of the model we also establish the conditions 
under which switch-like regulation of two fluxes is possible; from the 
extensive analysis we discover more than one possible mechanism of this 
phenomenon. 
\end{abstract}

\maketitle

\section{Introduction}

Totally asymmetric simple exclusion process (TASEP) 
\cite{schutz01,derrida98,mallick15} 
is one of the simplest models of a system of interacting self-propelled 
particles. In the simplest version of this model a fraction of the sites 
on a one-dimensional lattice are occupied by particles that can jump 
forward, at a given rate, if and only if its target site is not already 
occupied by another particle. For a finite lattice, boundary conditions 
are also specified. Under periodic boundary condition the lattice becomes, 
effectively, a closed ring and, therefore, no extra rate constants are 
required to define the kinetics completely. However, if open boundary 
condition is imposed the rates of entry and exit of the particles also 
have to be specified for a complete description of the kinetics of the model. 
TASEP sets a paradigm for nonequilibrium statistical mechanics of driven 
systems 
\cite{schutz01,derrida98,mallick15,mukamel00,blythe07}.

Various extensions of TASEP have been proposed to model vehicular 
traffic \cite{chowdhury00,schad10} where the sites of entry and exit of 
the particles are identified as the ON- and OFF-ramps, respectively, 
of the vehicles. Throughout this paper we refer to the segment of the 
lattice between the ON- and OFF- ramps as the ``track'' for the 
corresponding particles. TASEP has also been adapted to describe 
many traffic-like collective phenomena in biological systems 
(see ref.\cite{chowdhury05,chou11,chowdhury13a,rolland15} for reviews). 
For example, in TASEP of hard rods, each extended particle simultaneously 
covers ${\ell}$($>1$) successive lattice sites, instead of occupying just one 
site, thereby making those ${\ell}$ sites inaccessible to others although it 
can move forward by only one lattice site in each time step. This model was 
originally introduced to model protein synthesis \cite{macdonald68,macdonald69}. 
In this paper we introduce a new biologically motivated extension of TASEP 
and establish various possible mechanisms of an interesting phenomenon.

Single-species exclusion model has been extended in the past to multi-species 
particles (or rods) which move either in the same or in the opposite directions; 
in some of these extended models both species share the same track whereas 
in others distinguishable species of particles move along distinct tracks 
\cite{chowdhury00,schad10,chowdhury05,chou11,chowdhury13a,rolland15,macdonald68,macdonald69,tripathi08,klumpp08,klumpp11,sahoo11,ohta11,wang14,schutz03,kunwar06,john04,lin11,lakatos03,shaw03,shaw04a,shaw04b,chou03,chou04,zia11,chou99,levine04,liu10,ciandrini10,basu07,gccr09,greulich12,kuan15,chowdhury08,oriola15,sugden07,evans11,chai09,ebbinghaus09,ebbinghaus10,muhuri10,neri11,neri13a,neri13b,curatolo16}.  
In this paper we develop a class of novel biologically motivated exclusion models 
of two distinguishable species of hard rods, with their respective distinct pairs of  
ON- and OFF-ramps. 

In our models the positions of the rods on even two collinear tracks can be described 
by a single ``lattice'' of equidistance points, as we'll show in the next section. 
Consequently, in some of the models introduced in this paper the ON- and 
OFF-ramps do not necessarily coincide with the end points of the finite lattice.
So far as the direction of movement of the rods are concerned, we separately consider 
two different cases. In the first, both species of rods hop in the same direction  
although the lengths of tracks for the two species are different; therefore, all the 
nearest-neighbour encounters between the rods are {\it co-directional} irrespective 
of the species of rods involved in the encounter. In contrast, in the second case, 
the two species of rods hop in opposite direction; therefore, the intra-species 
encounters of the rods are still co-directional whereas the contra-directional hops 
of the two species of rods result in their head-on encounters. So far as the outcomes 
of the collision are concerned, we separately consider various possible (biologically 
motivated) scenarios; these include, for example, passing each other or premature 
detachment from the track.

One of the key quantities that characterize the non-equilibrium steady 
states of such driven systems is the current (or flux) of the particles 
that is defined as the number of particles passing through a lattice 
site per unit time. By a combination of mean-field theory and computer 
simulations of these theoretical models we calculate (a) the fluxes and  
(b) the spatial organization of the rods, both in the steady state of 
the system.  More specifically, we investigate how the fluxes and 
the density profiles of the two species of the rods depend on the 
(i) geometric parameters, like the relative orientation and spatial extent 
of the overlap of the two tracks, and (ii) kinetic parameters, like the 
rates of entry, exit, unhindered hopping as well as those of passing or 
premature detachments of the rods up on close encounter.

The results of our investigation elucidate the consequences of interference 
of the exclusion processes involving two distinguishable species of rods. 
For single-species TASEP under open boundary conditions the flux is 
determined by the interplay of its three rate constants, namely the 
rates of entry, exit and hopping in the bulk. However, in two-species 
exclusion models the flux of one species is expected to depend, in general, 
on the rate constants of the other. But, as we demonstrate here, 
the effect of one species on the other can be so strong that the flux 
of one can become vanishingly small when the flux of the other is high. 
In other words, we present ``proof of principle'' that a {\it switch-like} 
regulation of the flux of the two species of rods is possible in a two-species 
exclusion process. Most importantly, we establish the traffic conditions 
necessary for such regulation. We interpret the results physically. We also 
discuss the possible implications of the results in the context of the biological 
phenomenon that has motivated the formulation of these models.

\section{Model}
\label{sec-model}

\begin{figure}[t]
  \includegraphics[angle=0,width=0.9\columnwidth]{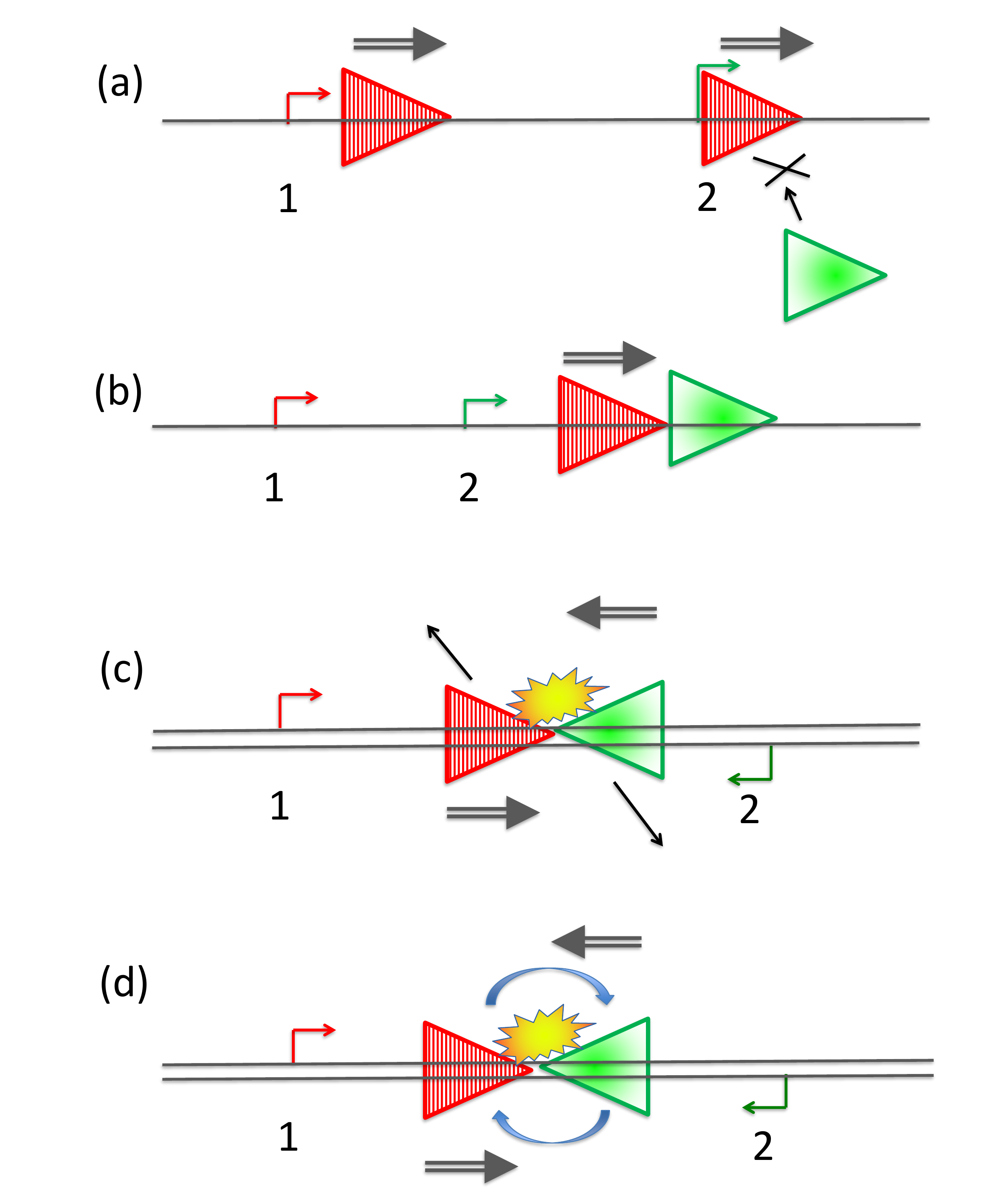} \\
  \caption{(Color online) Schematic representation of the common modes of Transcriptional Interference 
(TI): (a) occlusion (in co-directional TI), (b) road block (in co-directional TI), (c) head-on collision 
(in contra-directional TI) resulting is detachment, (d) head-on collision (in contra-directional 
TI) resulting in passing each other. Single bent arrows labelled by 1 and 2 indicate the sites of 
initiation of transcription of the genes 1 and 2. The two distinct species of RNAP are represented 
by shaded red and open green triangles where the orientation of the triangle indicates its 
natural direction of movement. Each double arrow indicates the direction in which the corresponding  RNAP has a natural tendency to move at that instant of time; absence of double arrow implies that the RNAP is 
stalled at that moment. The single straight arrows in (c) indicate the possible detachment resulting 
from head-on collision  whereas the semi-circular arrows in (d) depict the passing of the RNAPs 
while approaching each other head-on.
}
  \label{fig-TImodes}
\end{figure}


\begin{figure}[t]
(a)\\
  \includegraphics[angle=0,width=0.85\columnwidth]{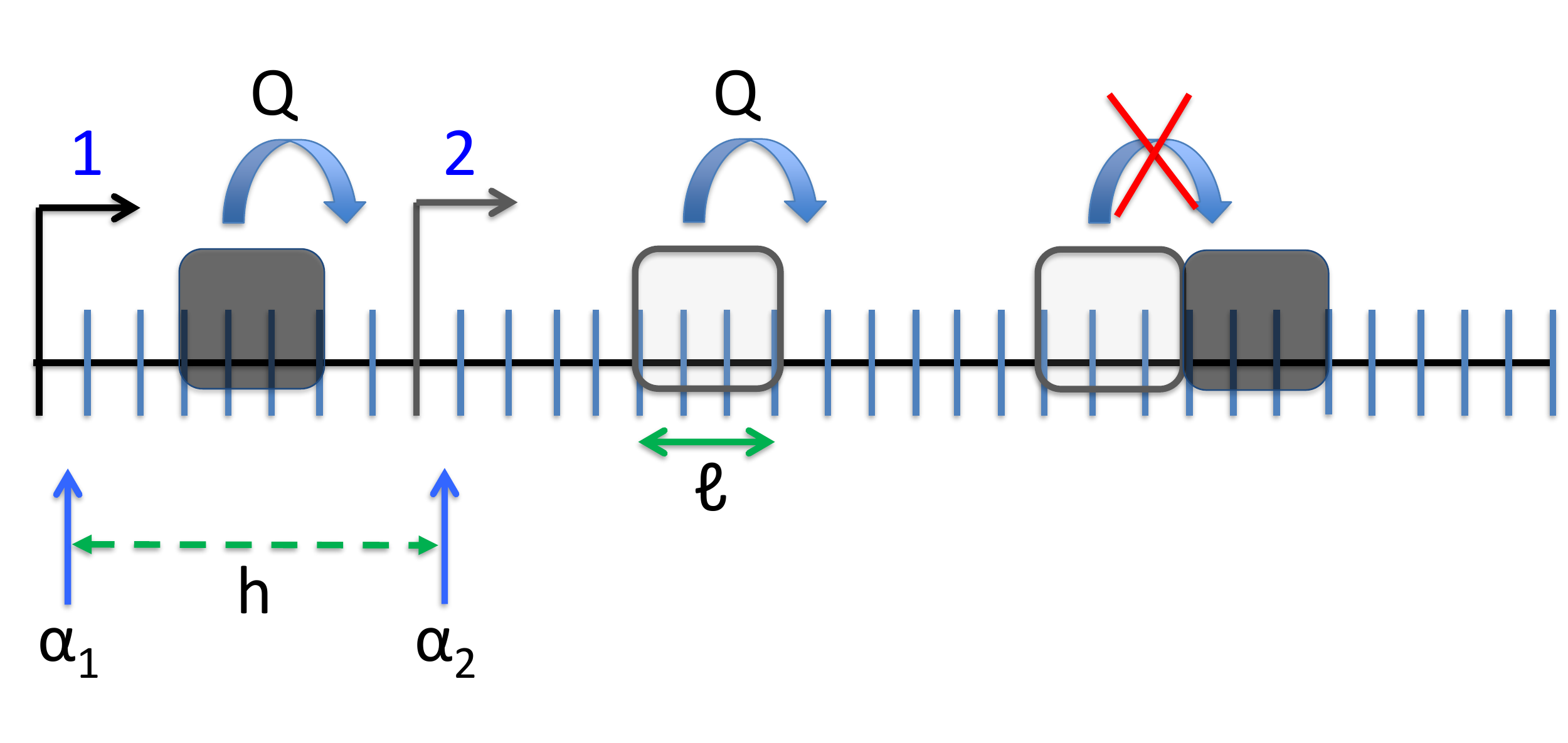} \\
(b) \\
  \includegraphics[angle=0,width=0.85\columnwidth]{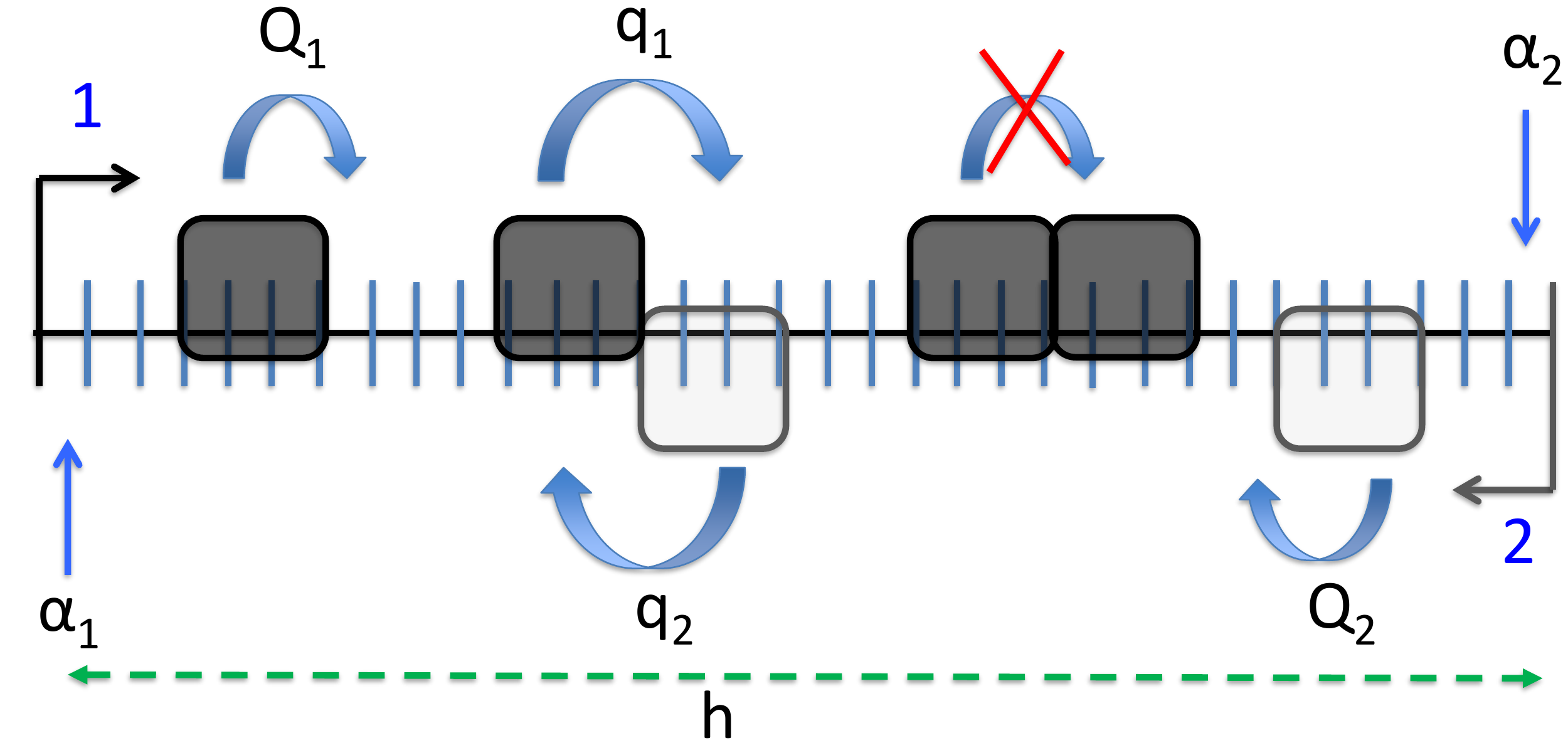} \\
(c) \\
  \includegraphics[angle=0,width=0.85\columnwidth]{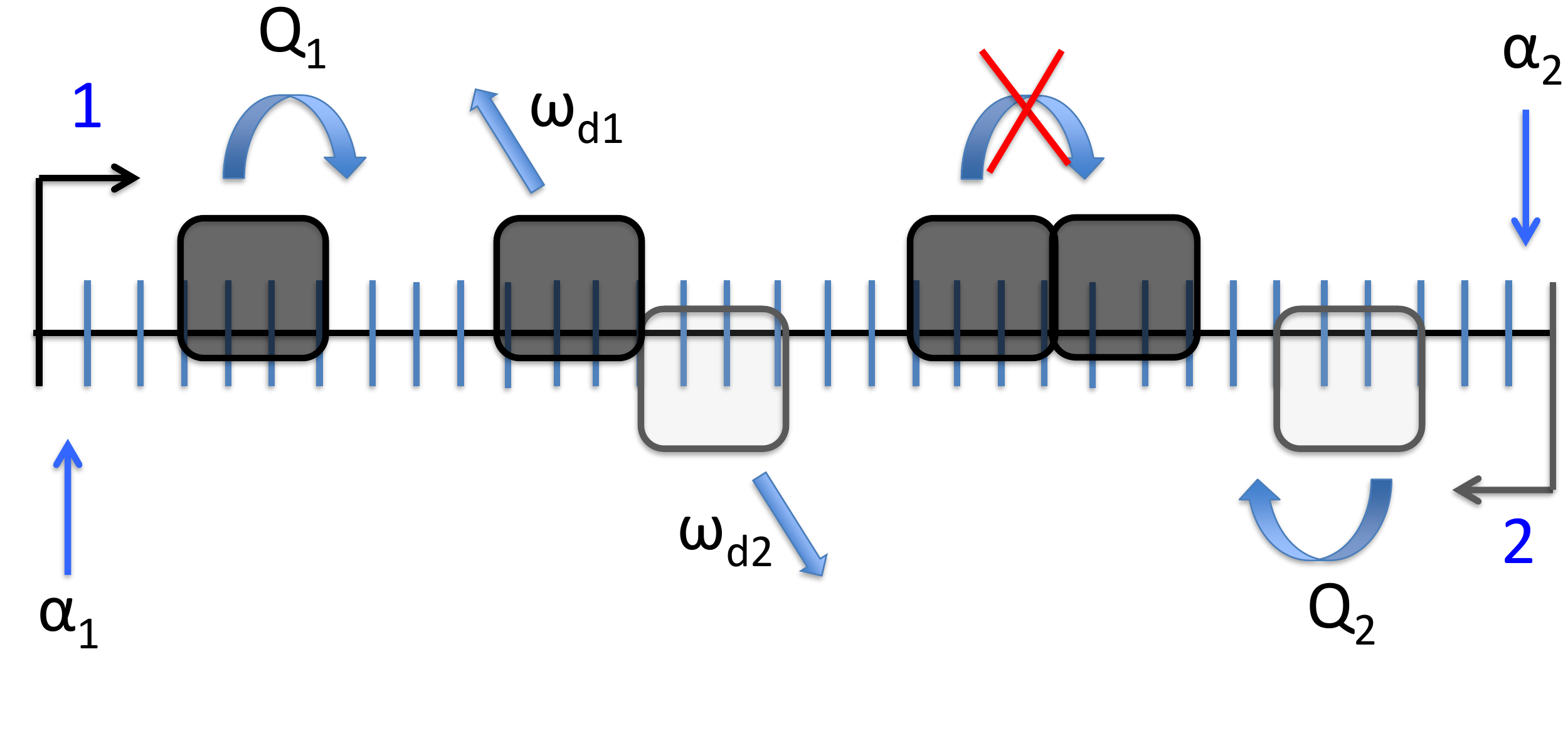} \\
  \caption{(Color online) Schematic representation of our two-species exclusion 
models that are motivated by Transcriptional Interference, which is a biological 
phenomenon sketched in Fig.\ref{fig-TImodes}. A single lattice is used for formulating 
the models for co-directional (in(a)) and contra-directional (in(b)-(c)) movement of the 
two species of rods.  The ON-ramps are marked by the bent arrows labelled by 
1 and 2; the OFF-ramps are also distinct (but not shown explicitly), in general. 
The two distinct species of rods are represented by filled and open rectangles. 
The symbol ${\ell}$ denotes the linear  size of each hard rod 
while the separation between the two ON-ramps is denoted by $h$. 
In all the cases (a)-(c) the members of the same species of rods use the 
same ON-ramp and carry the label of the ON-ramp (1 or 2) throughout their 
journey.
$\alpha_{1}$ and $\alpha_{2}$ are the rates of initiation (i.e., entry) of the two 
species of rods at their respective ON-ramps. The different hopping rates in the 
bulk under different conditions are  also shown along with the corresponding 
arrows (see the text for details). 
}
  \label{fig-model}
\end{figure}


\subsection{Biological motivation of the model: Transcriptional Interference (TI)}
\label{sec-biomotiv}

The models developed in this paper are primarily motivated by a specific 
type of traffic-like collective phenomena in living cells. 
Synthesis of messenger RNA, a heteropolymer, using the corresponding 
template DNA, is called transcription; it is carried out by a molecular 
machine called RNA polymerase (RNAP) \cite{buc}.  
This machine also exploits the DNA template as a filamentous track for 
its motor-like movement consuming input chemical energy \cite{buc}.
Polymerization of each RNA by a RNAP takes place normally in three stages: 
(a) initiation at a specific `start' site (also called initiation site) 
on the template, (b) step-by-step elongation of the RNA, by one nucleotide 
in each forward step of the RNAP motor, and (c) termination at a specific 
`stop' site (also called termination site) on the template.  
For the sake of convenience, throughout this paper we refer to the segment 
of the template DNA between the start and the stop sites as a `gene'.
A RNAP locally unzips the double stranded DNA creating a ``bubble'' thereby 
accessing a single DNA strand that serves as the template. The RNAP and 
the DNA bubble, together with the growing RNA transcript, forms a transcription 
elongation complex (TEC).

Often multiple RNAPs transcribe the same gene simultaneously. 
In such RNAP traffic \cite{chowdhury05,chou11}, all the RNAPs engaged 
simultaneously in the transcription process move in the same direction 
while polymerizing identical copies of a RNA, all by initiating 
transcription from the same start site and, normally, terminating at the 
same stop site.  Any segment of the template DNA covered by one RNAP 
(more appropriately, covered by a TEC) is not accessible simultaneously 
to any other RNAP. Moreover, a typical RNAP covers several nucleotides. 

In the TASEP-based models of RNAP traffic on the template DNA strand 
the one-dimensional lattice represents a single-stranded DNA (ssDNA). 
Each site of this lattice denotes a nucleotide which is a monomeric 
subunit of the DNA track. Each RNAP (more appropriately, each TEC) 
is represented by a hard rod that simultaneously covers more than one 
site of the lattice. Although each TEC is slightly larger than the RNAP, 
the distinction is ignored in most of the TASEP-based models of transcription. 
We also ignore that slight difference in our model and use the term RNAP 
throughout instead of the more appropriate term TEC. The TASEP-based 
models of transcription reported so far also do not take into account the 
detailed structure of initiation site (for example, the so-called promoter), 
do not explicitly include the molecules that assist in the process of transcription 
initiation by the RNAP and capture the multi-step bio-chemical kinetics by a 
single rate constant. The sites of 
initiation and termination of transcription on the template are 
represented by the ON- and OFF-ramps, respectively, on the track for 
the rods. Single-site stepping rule is motivated by the fact that a 
RNAP must transcribe the successive nucleotides one by one. The steric 
exclusion between the RNAPs is naturally captured by TASEP-type models 
for RNAP traffic \cite{tripathi08,klumpp08,klumpp11,sahoo11,ohta11,wang14} 
because each RNAP is represented by a rigid rod. 

The theoretical models developed in this paper are motivated by more complex 
RNAP-traffic phenomena, called transcriptional interference (TI) 
\cite{shearwin05,mazo07} that are believed to play important regulatory roles 
in living cells \cite{pelechano13,georg11,lapidot06,kornienko13}.  
These phenomena arise from simultaneous transcription of two overlapping 
genes either on the same DNA template or two genes on the two adjacent 
single strands of a duplex (double-stranded) DNA (see Fig.\ref{fig-TImodes}). 
In the former case traffic is entirely uni-directional, as sketched in  
Figs.\ref{fig-TImodes}(a) and (b), although RNAPs transcribing different 
genes polymerize two distinct species of RNA molecules by starting (and 
stopping) at different sites on the same template DNA strand. 
In the case of co-directional TI both the genes may share a common 
termination site (off-ramp); that situation is captured as a special case 
of our more general formulation. 

In the case of contra-directional TI, as sketched in Figs.\ref{fig-TImodes}(c) 
and (d), RNAP traffic in the two adjacent ``lanes'' move in opposite directions 
transcribing the respective distinct genes. If the sites of termination of both the 
transcriptional processes are outside the region of the overlap of the two genes 
encoded on the two adjacent DNA strands the arrangement is defined as `
`head-to-head''. In contrast, in the ``tail-to-tail'' arrangement of the genes the 
sites of initiation marked on the two adjacent strands of the duplex DNA are 
beyond the region of overlap of the two genes.
In both these situations the two interfering transcriptional processes have 
suppressive effects on each other \cite{shearwin05,mazo07}.

In general, a RNAP at the initiation, elongation or termination stage
of transcription of one gene can affect the initiation, or elongation 
(or induce premature termination) of that of the other  gene by another 
RNAP \cite{pelechano13,georg11,lapidot06}. 
In other words, the stages of transcription of the two interfering RNAPs 
define a distinct mode of interference. Different modes of interference 
have been assigned different names like ``occlusion'' (Fig.\ref{fig-TImodes}(a)), 
``road block'' (Fig.\ref{fig-TImodes}(b)), ``collision'' Figs.\ref{fig-TImodes}(c) and (d), 
etc \cite{shearwin05, sneppen05}.

\subsection{Model of two-species exclusion process motivated by TI }
\label{sec-TaI}

We label the two species of rods by the integer indices 1 and 2. For the  
convenience of mathematical formulation of the model, we use a single 
lattice, with the lattice sites marked by the integer index $i$ ($i=1,2,...$ 
from left to right) to denote the positions of both species of rods. In 
general, the distance from the respective ON-ramps to the corresponding 
OFF-ramps for the two species can be different (see Fig.\ref{fig-model}). 
As stated in the introduction, the segment of the lattice between the ON- 
and OFF-ramps is defined as the {\it track} for the corresponding species 
of rods; $L_1$ and $L_2$ denote the lengths of the two tracks, measured 
in terms of the number of lattice sites, for the species 1 and 2, respectively. 
We identify two sites separated by $h$ sites as the ON-ramps for the two 
species; $h$ is a non-negative integer; negative $h$ essentially corresponds 
to interchanging the labels 1 and 2 (see fig.\ref{fig-model}). 
Any change of $h$, keeping both $L_{1}$ and $L_{2}$ fixed, alters the 
extent of overlap of the two tracks. 

The length of each hard rod is ${\ell}$ in the units of lattice sites, 
i.e., it covers ${\ell}$ successive sites of the lattice simultaneously 
thereby making these sites inaccessible to any other rod. 
We denote the position of a rod of species 1 by the lattice site at which 
the {\it leftmost} unit (the left edge) of the rod is located. In the 
terminology used consistently throughout this paper, the site $i$ where the 
the leftmost unit of the rod of species 1 is located is said to be ``{\it occupied}'' 
by the rod while the next ${\ell}-1$ sites of the lattice are merely 
``{\it covered}'' by the same rod. Thus, if the lattice site $i$ denotes the position 
of a rod on track 1 then the rod covers not only the site $i$ but also 
the next ${\ell}-1$ sites $i+1, i+2,...,i+{\ell}-1$. But, in case of 
contra-directional movement of the two species the position of a rod of 
species 2 is denoted by the lattice site $i$ at which its {\it rightmost} unit 
(the right edge) is located. This site is said to be ``occupied'' by the rod 
of species 2 while the rod ``covers''  the ${\ell}$ sites $i, i-1, i-2,...i-{\ell}+1$. 
The rods interact with each other with only hard core repulsion that is captured 
by imposing the condition that no site on a given lattice is allowed to be {\it covered} 
by more than one rod simultaneously. For reasons which will become clear 
when we present our results, the longer is the ${\ell}$ the stronger is the 
suppressive effect of one transcriptional process, i.e., one TASEP, on the other.

The entry of a rod, however, is not possible as long as one or more 
of the first ${\ell}$ sites, starting from the ON-ramp marked on that 
track, remain covered by any other rod, irrespective of the identity of 
the latter. We denote the rates of entry of the rods of species 1 and 2 on the 
corresponding tracks by $\alpha_1,\alpha_2$, respectively.
Whenever first ${\ell}$ successive sites, starting from 
the ON-ramp on a track is vacant, a fresh rod is allowed to cover those 
${\ell}$ sites thereby indicating entry of that rod. Upon its entry, 
each rod receives an unique label $1$ or $2$ depending on which of the 
two ON-ramps through which it enters; accordingly it belongs to the species  
1 or 2 and it carries its label throughout its journey along the 
corresponding track. Irrespective of the actual numerical value of 
${\ell}$, each rod can move forward by only one site in each step,
provided the target site is not already covered by any other rod. 
Unless prematurely detached from its track under special situations that 
we discuss in section \ref{sec-contradir}, a rod  would detach from its 
track  after it reaches the OFF-ramp of the corresponding track. So far 
as the rates of the detachment of a rod from the exit site is concerned, 
we denote the corresponding rates by $\beta_1$ and $\beta_2$, respectively.  

The rods and the tracks in our  two-species exclusion model  correspond 
to the RNAPs (more appropriately TECs) and genes in transcriptional 
interference.  Initiation and termination of transcription are captured by the 
entry and exit (at the designated ON- and OFF-ramps), respectively, of the 
rods onto their respective tracks. For obvious reason, in analogy with 
transcriptional interference, we call this general phenomenon as {\it TASEP 
interference} (TaI).
Interference of co-directional transcriptions of two overlapping genes 
encoded on the same DNA strand is quite well known \cite{shearwin05}; 
the TaI model in Fig.\ref{fig-model}(a) captures the most essential 
kinetic aspects of this process. In the context of interference of 
contra-directional transcription of two geometrically overlapping 
genes encoded on the two adjacent strands of a duplex DNA, passing of 
bacteriophage (viruses that invade bacteria) RNAPs, without premature 
detachments, has been observed experimentally \cite{ma09}. This scenario 
would correspond to the model depicted schematically in Fig.\ref{fig-model}(b). 
On the other hand, premature detachment of RNAPs, instead of passing, 
is more common \cite{shearwin05}, particularly among non-bacteriophage 
RNAPs. This situation is captured in our TASEP-based model shown in 
Fig.\ref{fig-model}(c).

In most of the earlier theoretical models on RNAP traffic 
\cite{tripathi08,klumpp08,klumpp11,sahoo11,ohta11,wang14} 
all the RNAPs were engaged in transcribing a single gene; therefore the 
traffic was uni-directional and every RNAP polymerized identical copies 
of the RNA while a single pair of start-stop sites marked the points of 
initiation and termination of transcription. In contrast, in the models 
of TaI reported in this paper two distinct pairs of ON- and OFF-ramps 
correspond to the two distinct pairs of start-stop sites for the 
initiation and termination of the respective genes. 
We have ignored the possibility of backtracking of 
the individual rod which is a well known phenomenon for RNAPs engaged 
in transcription \cite{sahoo11}. In future extensions of our model 
\cite{ghosh15} we intend to explore the effects of backtracking 
\cite{sahoo11} as well as active re-starting of stalled rod by a trailing 
rod \cite{dong12}.

In all the earlier theoretical works on transcriptional interference the 
effects of the different modes of interference, e.g., occlusion, 
road block, collision, etc., (shown schematically in Fig.\ref{fig-TImodes}) 
have been studied separately \cite{sneppen05,palmer09}. 
The simple unified model of TaI that we have developed here not only 
captures all possible modes of transcriptional interference but also 
accounts for the co-directional and contra-directional traffic of both 
bacteriophage RNAPs and non-bacteriophage RNAPs as various special cases.

\subsection{Quantities of interest and methods of calculation}

Let $P_{\mu}(j,t)$ denote the probability that at time $t$ the site $j$ 
is occupied by a rod of the $\mu$-th species ($\mu=1,2$ 
for the species 1 and 2, respectively), as per the convention adopted 
earlier in this section. In the steady state $P_{\mu}(j,t)$ are 
independent of time. The number of rods passing through an arbitrary site 
$j$ per unit time is defined as the {\it flux} of rods at that site. Therefore,
the flux of the two species of rods measured at an arbitrary site $j$ is given by 
\begin{equation}
J_{\mu}(j) = \kappa P_{\mu}(j) \times {\rm Prob.~ that~ target~ site~ is ~not ~covered}
\label{eq-Jprofile}
\end{equation}
where $\kappa$, the allowed hopping rate at $j$, could be $Q$, or 
$q_1$ or $q_2$ or $\beta_1$ or $\beta_2$ depending on the position of 
the site $j$ on the lattice and the relative directions of movements 
of the rods, etc. as described above. Thus, $J_{\mu}(j)$ gives the 
flux profile along the lattice. In the absence of premature detachments, 
the steady state flux of the rods is independent of the site $j$, i.e., the 
flux profile is flat. But, if premature detachment of the rods, induced by 
head-on collisions, is allowed, the flux of the rods along the track would 
decrease with increasing distance from the corresponding ON-ramp. 
The flux of the rods of the two species at the respective OFF-ramps 
are true measures of the corresponding overall rates of successful 
transcription events (i.e., synthesis of full length RNA transcripts); 
these are obtained from 
\begin{equation}
J_{1} = \beta_{1} P_{1}(L_1),~~{\rm and}~~ J_{2} = \beta_{2} P_{2}(L_2).
\label{eq-Fdefn}
\end{equation}  
where exit from the OFF-ramp does not require accessibility of any target 
site.

The master equations governing the time evolution of $P_{\mu}(j,t)$
are set up in the next subsection. 
Since these master equations could not be solved analytically, we 
solved these numerically to obtain the steady-state solutions. 
In the steady state the mean-field equations reduce to a set of large 
number of coupled algebraic equations for $P_{\mu}(j)$, two for each 
lattice site $j$ corresponding to $\mu=1,2$. Solving these equations 
self-consistently one can, in principle, get the steady state solutions 
$P_{\mu}(j)$ ($j=1,2,...$). However, in our numerical approach, we 
treated the master equations as a set of coupled ordinary differential 
equations. Starting from a suitably chosen initial profile $P_{\mu}(j,0)$, 
all the master equations for $P_{\mu}(j,t)$ were integrated with respect 
to time  iteratively in steps of infinitesimal time durations so as to 
converge to the steady-state values $P_{\mu}(j)$. 
Using the two numerical values of $P_{1}(L_1)$ and $P_{2}(L_2)$ in 
(\ref{eq-Fdefn})we get the steady-state fluxes $J_{1}$ and $J_{2}$, 
respectively, at the corresponding OFF-ramps. 
The plot of the steady-state values of $P_{\mu}(j)$ against the site index 
$j$ would give the profile of the number density of the corresponding 
species ($\mu=1,2$) of RNAPs in the steady state. 

In order to test the range of validity of the MFA made in writing the 
master equations, we also carried out extensive direct computer 
simulations (Monte Carlo simulations) of our model using the same set 
of parameter values that we used for solving the master equations. 
Starting from an initial condition with empty lattices, the system was updated 
using a random-sequential algorithm \cite{schad10}
for sufficiently large number of Monte Carlo steps (typically, 
two million), while monitoring the flux, to ensure that it reached the 
steady state. 

The steady state data were collected over the next five million  
Monte Carlo steps. In the computer simulations $J_{1}(j)$ and $J_{2}(j)$ were 
obtained by counting the number of RNAPs passing through the site $j$  
per Monte Carlo step. The steady-state flux as well as the density profiles of 
the rods presented in this paper are averages of the data collected only 
in the steady state of the system. Finally, these data were averaged over 
many runs each starting from the empty-lattice initial states. Unless 
stated otherwise, the symbols $J_{1}$ and $J_{2}$ would refer to 
$J_{1}(L_1)$ and $J_2(L_2)$, respectively, i.e., the steady-state fluxes 
measured at the OFF-ramps of the corresponding species of rods.

The mean-field theoretical predictions on the steady state values of 
the density profiles and those of the fluxes $J_1(j)$ and $J_2(j)$ are 
plotted in the figures using {\it continuous, dashed- and dotted lines}.  
The discrete symbols triangle ($\blacktriangle$), dot ($\bullet$), etc. 
are used in the figures to present the numerical data for the density 
profiles and fluxes, obtained from our computer simulations.

All the numerical results plotted in this paper have been obtained for 
the numerical values of the parameters listed in table \ref{tab-parameters}; 
parameters not listed in this table were varied over different ranges for 
graphical plots. By comparing with the results 
for a few other values of ${\ell}$, $L_1$ and $L_2$, we ensured that 
our conclusions do not suffer from any artefacts of the choice of these 
parameters. 

\begin{table}
\begin{tabular}{|c|c|c|c|} \hline
  $\ell$& $Q_{1}=Q_{2}=Q$ &$q_{1}=q_{2}$&$\omega_{d1}=\omega_{d2}$ \\\hline
  10 & 1  &0.33&0.33\\\hline
\end{tabular}
\caption{Numerical values of some model parameters.}
\label{tab-parameters}
\end{table}

We emphasize that the main questions addressed in 
most of the earlier TASEP-based models are also fundamentally different 
from those addressed  in this paper. The main emphasis here is on the 
nature of {\it regulation} of flux of one species of rods by controlling the level 
of flow of another through TaI and correlating the variation of the flux with 
that of the density profiles.

\section{Co-directional traffic}
\label{sec-codir}

\begin{figure}[ht]
(a) \\
    \includegraphics[width=0.8\columnwidth]{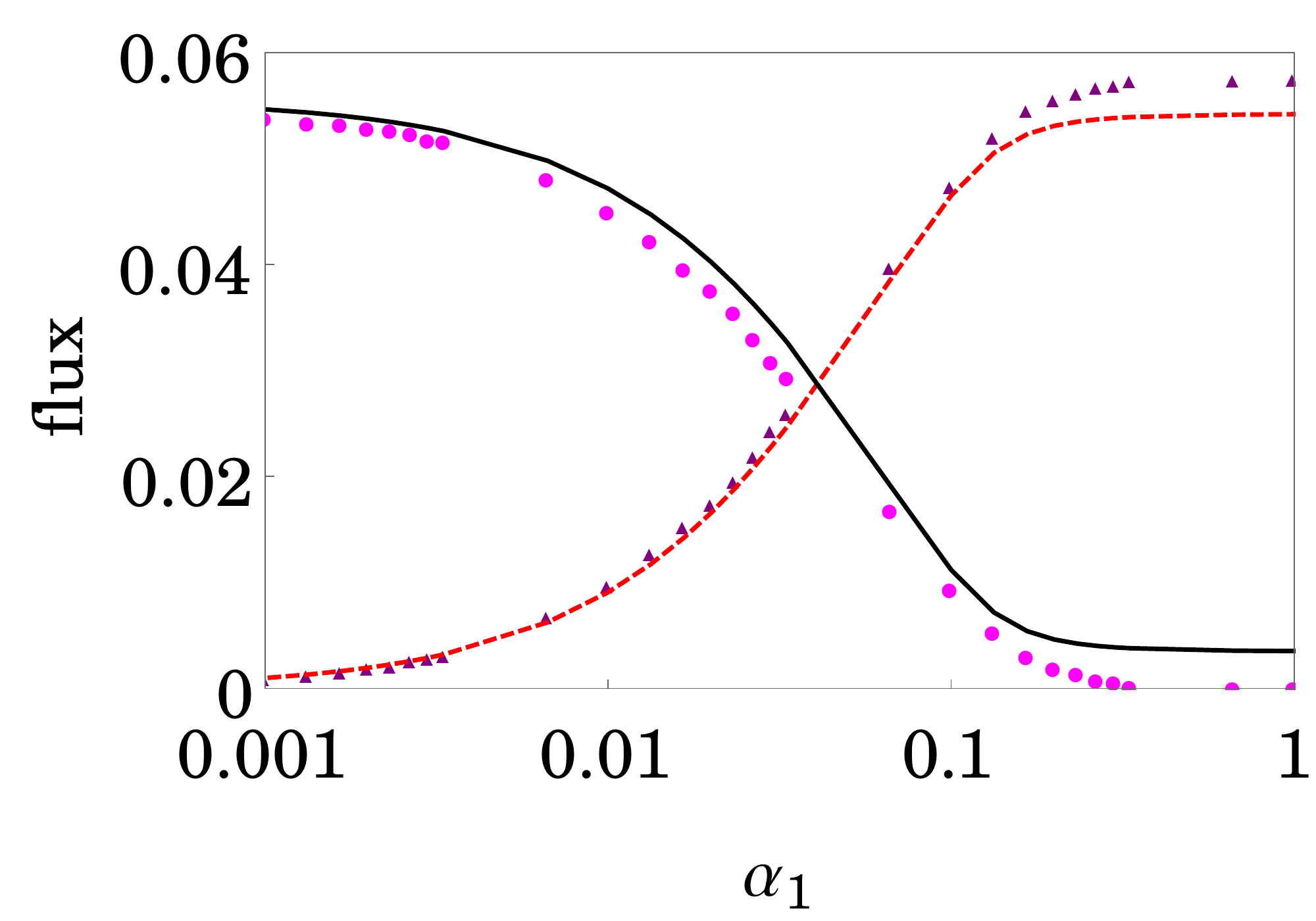}\\
(b)\\
    \includegraphics[width=0.8\columnwidth]{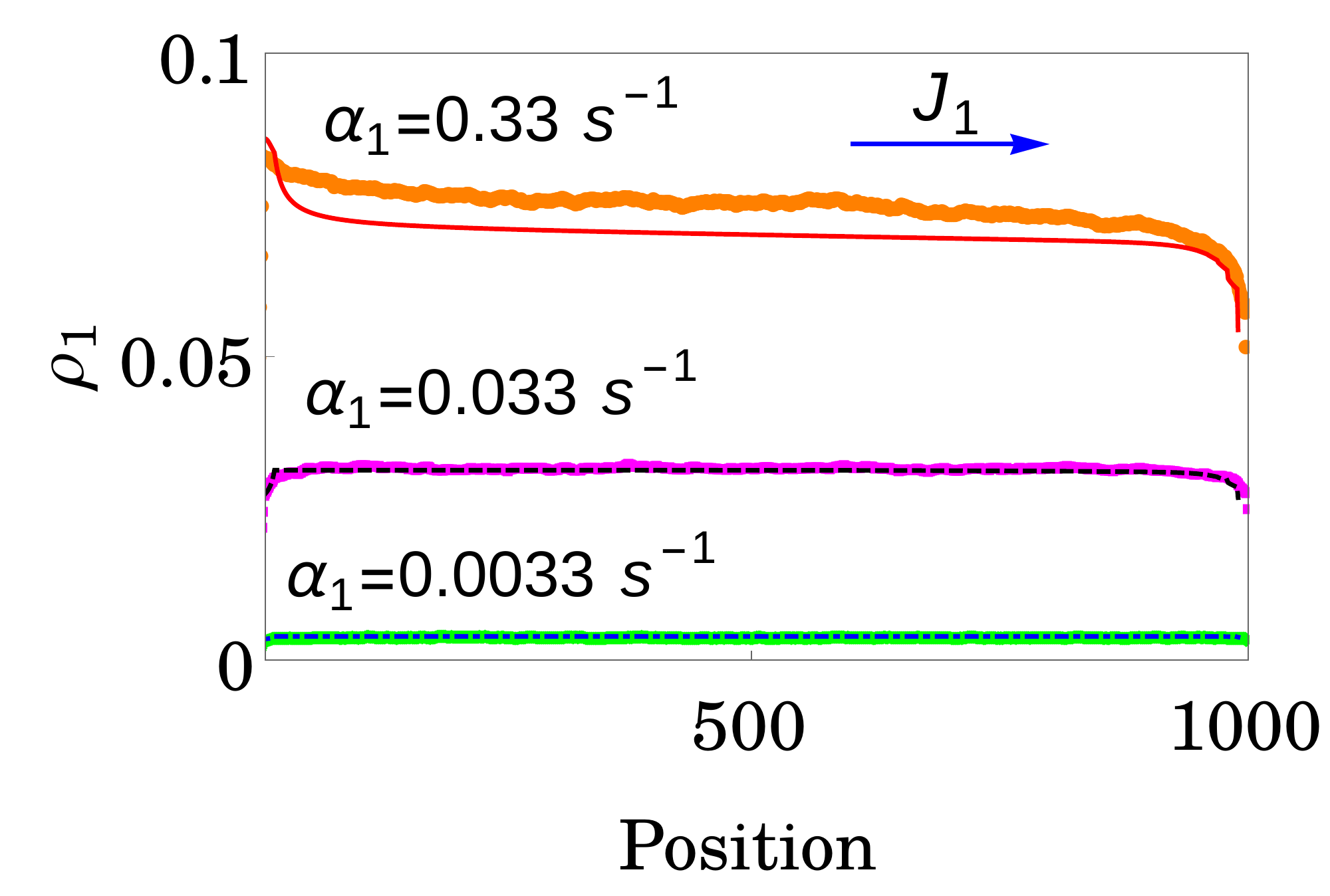}\\
(c)\\
    \includegraphics[width=0.8\columnwidth]{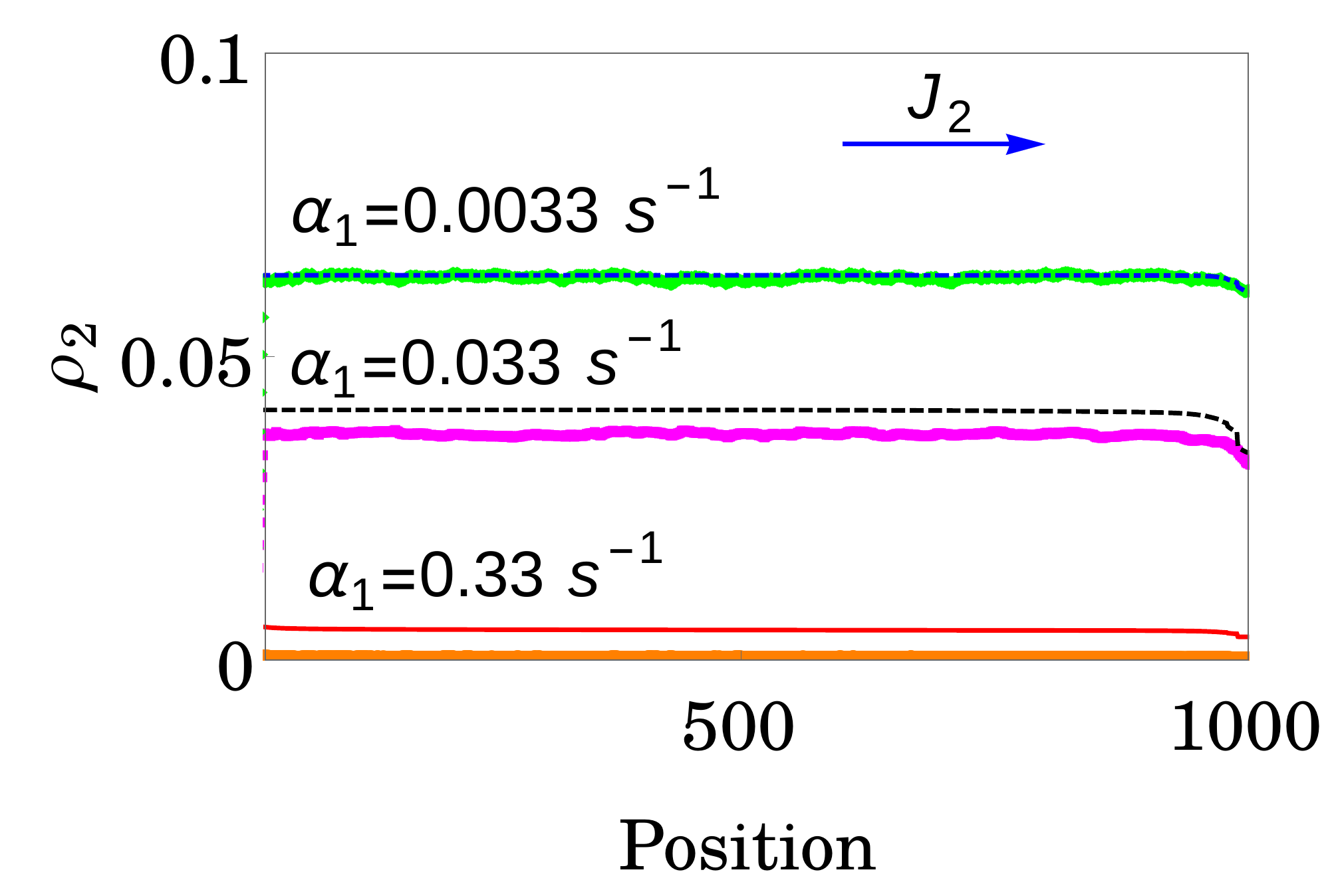}\\
  \caption{(Color online) Codirectional TaI: (a) The fluxes $J_{1}$ and $J_{2}$  
are plotted as functions of ${\alpha}_{1}$, for ${\alpha}_{2}=0.167$, $\beta_1 = \beta_2 = 1.0$ 
and $h=20$. The dashed and continuous lines are used to plot the mean field predictions 
for $J_{1}$ and $J_{2}$, respectively, while the corresponding discrete data obtained from 
computer simulations have been plotted using triangles and filled circles. Increasing $\alpha_{1}$ 
leads to switching OFF the TASEP 2 and switching ON the TASEP 1. The average density profiles 
$\rho_{1}$ and $\rho_{2}$, for three different values of ${\alpha}_{1}$ are plotted in (b) and (c), 
respectively; the lines and discrete data points correspond to  mean-field theory and computer 
simulations.
}
  \label{fig-switchCO}
\end{figure}

In our model of co-directional TaI (see Fig.\ref{fig-model}(a)) no rod can pass the 
other immediately in front of it irrespective of the whether the rod in front belongs 
to species 1 or 2. This is motivated by the fact that in case of co-directional 
TI both species of RNAPs move on the same single stranded DNA 
and, therefore, passing is not possible. For simplicity, we consider symmetric 
case so that both species of rods can jump forward with the same rate $Q$ 
if there is no obstruction in front of it.

The probability that the site $i$ is occupied by the left edge of a rod, 
irrespective of whether it is of type 1 or type 2, is given by 
$P(i) = \sum_{\mu=1}^{2} P_{\mu}(i)$.  Let $P(\underline{i}|j)$ be the 
conditional probability that, given that the site $i$ is occupied by (by definition, 
the left edge of) a rod, the downstream site $j$ is also occupied by (the left edge 
of) another rod. Obviously, $\xi(\underline{i}|j) = 1 - P(\underline{i}|j)$ 
is the conditional probability that, given that the site $i$ is occupied by a rod, the 
site $j$ is empty (i.e., not even covered by any rod). Under mean-field approximation 
this conditional probability reduces to the simple form \cite{shaw04a}
\begin{eqnarray}
\xi(\underline{i}|i+{\ell})=\frac{1-\sum\limits_{s=1}^{{\ell}} P(i+s)}{1+ P(i+{\ell}) - \sum\limits_{s=1}^{{\ell}} P(i+s)} 
\end{eqnarray}\\
Let $\xi(i)$ be the probability that site $i$ is not {\it covered} by any 
rod, irrespective of the state of occupation of any other site; by  
definition,  
\begin{equation}
\xi(i) = 1-\sum_{s=0}^{{\ell}-1}P(i-s).
\label{eq-xiDEF}
\end{equation}
Note that, if site $i$ is given to be occupied by the left edge of one rod, 
the site $i-1$ can be covered by another rod if, and only if, the site $i-{\ell}$ 
is also occupied  by the left edge of another rod.

\begin{figure}[ht]
(a) \\
    \includegraphics[width=0.85\columnwidth]{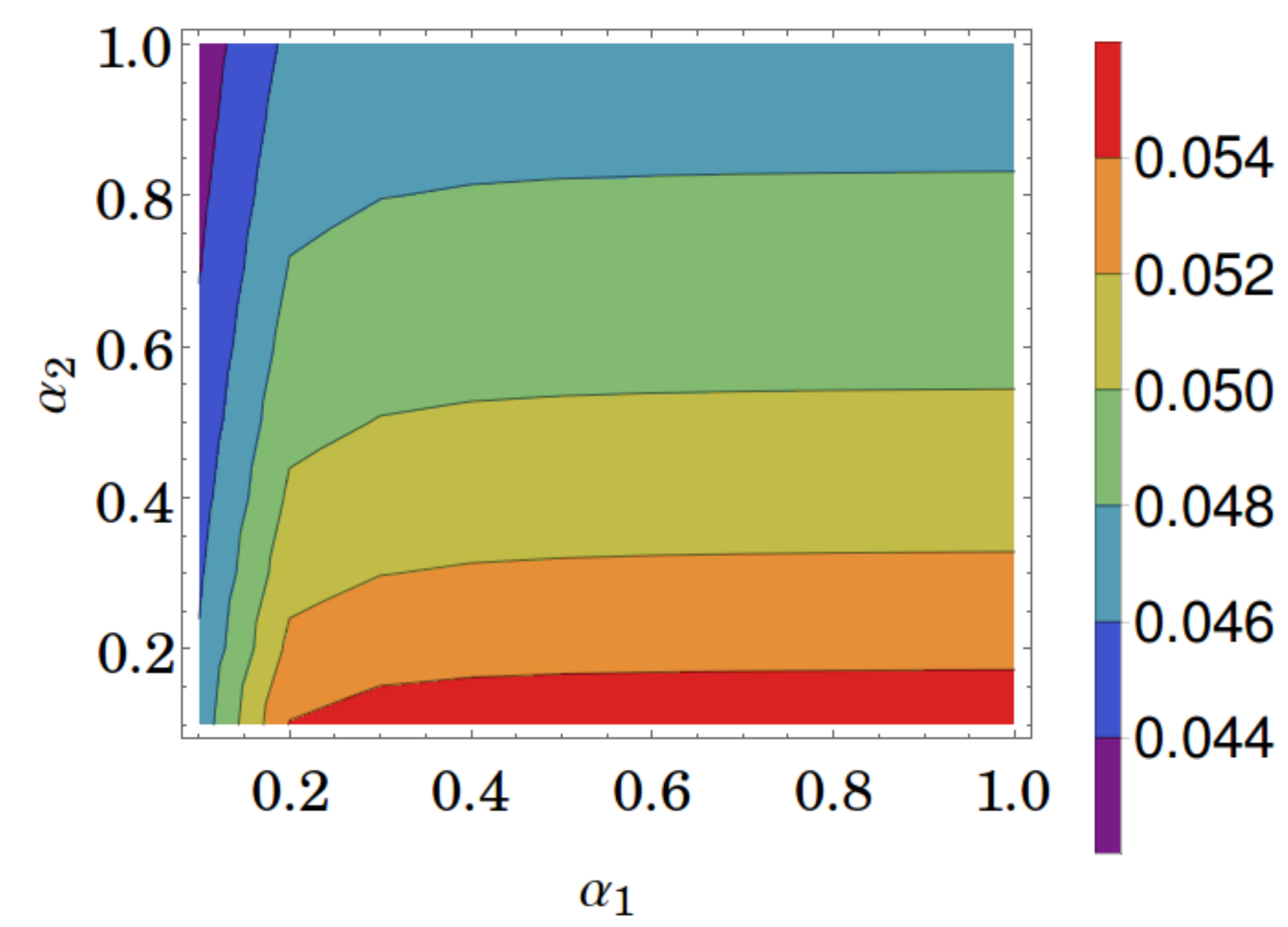}\\
(b)\\
    \includegraphics[width=0.85\columnwidth]{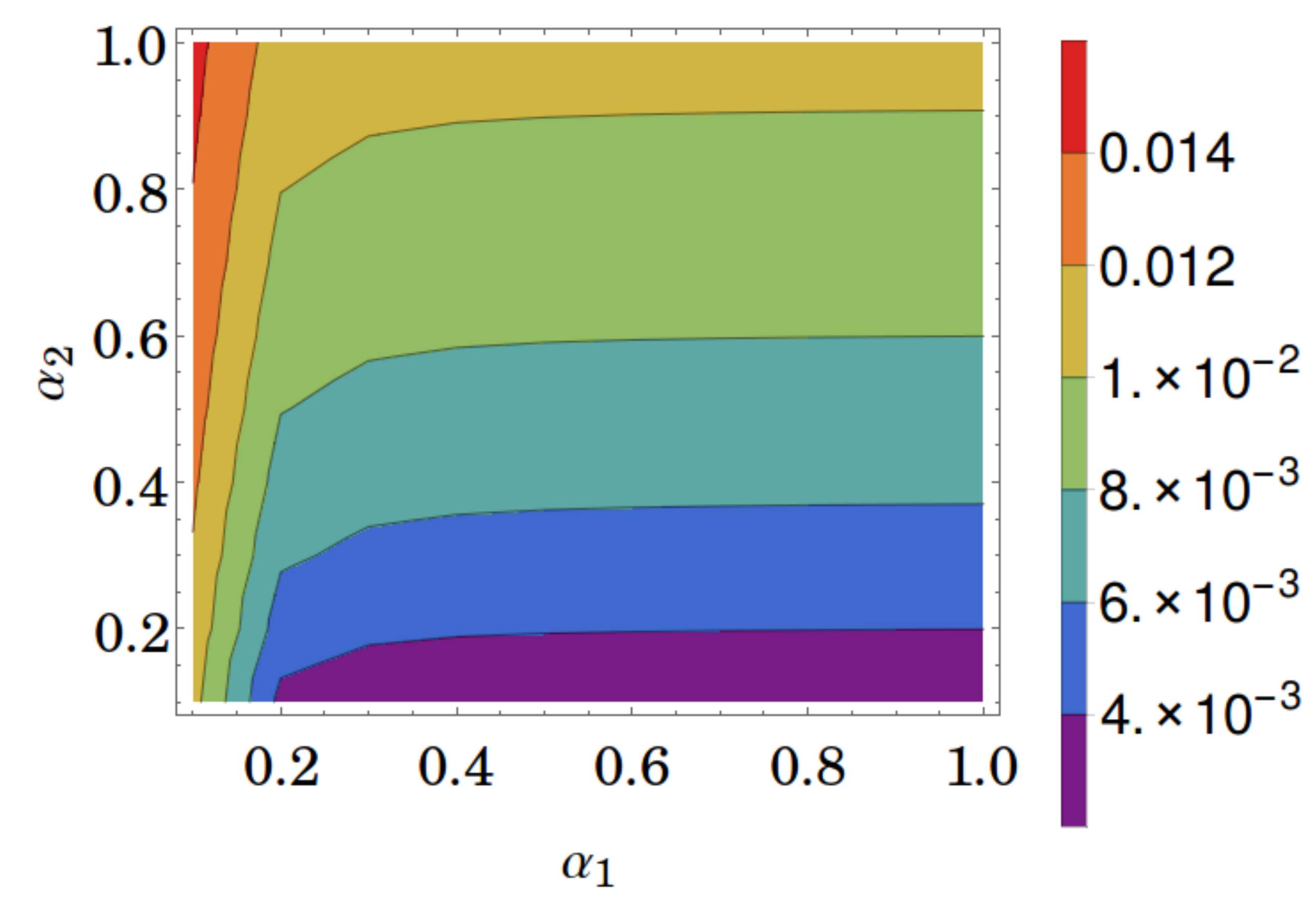}\\
  \caption{(Color online) Codirectional TaI: Contour plots of (a) $J_{1}$ and (b) $J_{2}$ in the 
$\alpha_{1}-\alpha_{2}$ plane obtained from mean-field theory for $\beta_{1}=\beta_{2}=\beta = 1.0$. Contours of constant flux is painted by the same color. 
}
  \label{fig-ContourCOdir}
\end{figure}

The master equations governing the stochastic kinetics of the two {\it species of rods} are given by 
\begin{widetext}
  \begin{eqnarray}
    \frac{dP_{1}(1,t)}{dt} &=& {\underbrace{~\alpha_1 \Biggl(1-\sum_{s=1}^{{\ell}}~P(s)\Biggr)}_{\text {Entry at ON-ramp}}}
    - {\underbrace{~Q P_{1}(1,t) ~\xi(\underline{1}|1+{\ell})}_{\text {Forward Hopping from j=1}}}~,~\nonumber\\
    \frac{dP_{1}(i,t)}{dt} &=&  {\underbrace{Q P_{1}(i-1,t)\xi(\underline{i-1}|i-1+{\ell})}_{\text {Forward Hopping to j=i}}}   -  {\underbrace{Q P_1(i,t) \xi(\underline{i}|i+{\ell})}_{\text {Forward Hopping from j=i}}}~~~{\rm ~for~}, ~(1<i<L_1)~, \nonumber \\
    \frac{dP_{1}(L_1,t)}{dt} &=&  {\underbrace{~Q P_{1}(L_1-1,t)\xi(\underline{L_1-1}|L_1-1+{\ell})}_{\text {Forward Hopping to j=$L_1$}}} -  {\underbrace{\beta P_1(L_1,t)}_{\text {Exit at OFF-ramp}}}~.\nonumber\\
        \label{eqs:1}
    \end{eqnarray}

  \begin{eqnarray}
    \frac{dP_{2}(1+h,t)}{dt} &=& ~\alpha_2 \xi(1+h)  \Biggl(1-\sum_{s=1}^{{\ell}} P(s+h)\Biggr) - Q P_{2}(1+h,t) ~\xi(\underline{1+h}|1+h+{\ell})~, \nonumber \\
    \frac{dP_{2}(i,t)}{dt} &=&  ~Q P_{2}(i-1,t)\xi(\underline{i-1}|i-1+{\ell}) ~ - ~ Q P_2(i,t) \xi(\underline{i}|i+{\ell})~~{\rm~ ~for~ },~ (1+h<i<L_2+h)~, \nonumber \\
    \frac{dP_{2}(L_2+h,t)}{dt} &=&  ~Q P_{2}(L_2+h-1,t)\xi(\underline{L_2+h-1}|L_2+h-1+{\ell}) - \beta P_2(L_2+h,t)~.\nonumber\\
    \label{eqs:2}
\end{eqnarray}
\end{widetext} 
where, from (\ref{eq-xiDEF}), $\xi(1+h)= 1 - [P(1+h)+P(h)+P(h-1)+\cdot \cdot \cdot+P(1+h-({\ell}-1))]$ 
accounts for the exclusion of the entry of a type 2 rod by the rods of type 1 (i.e., occlusion of 
the site of initiation of transcription of gene 2 by the RNAPs engaged in the transcription of gene 1).
Obviously, $\xi(1+h) = 1$ for $h=0$, i.e., if both species of rods use the same ON-ramp.
Replacement of the conditional probabilities in these equations by the 
expressions derived above in terms of the single-site probabilities is 
equivalent to MFA.

\begin{figure}[htb]
\includegraphics[angle=0,width=0.9\columnwidth]{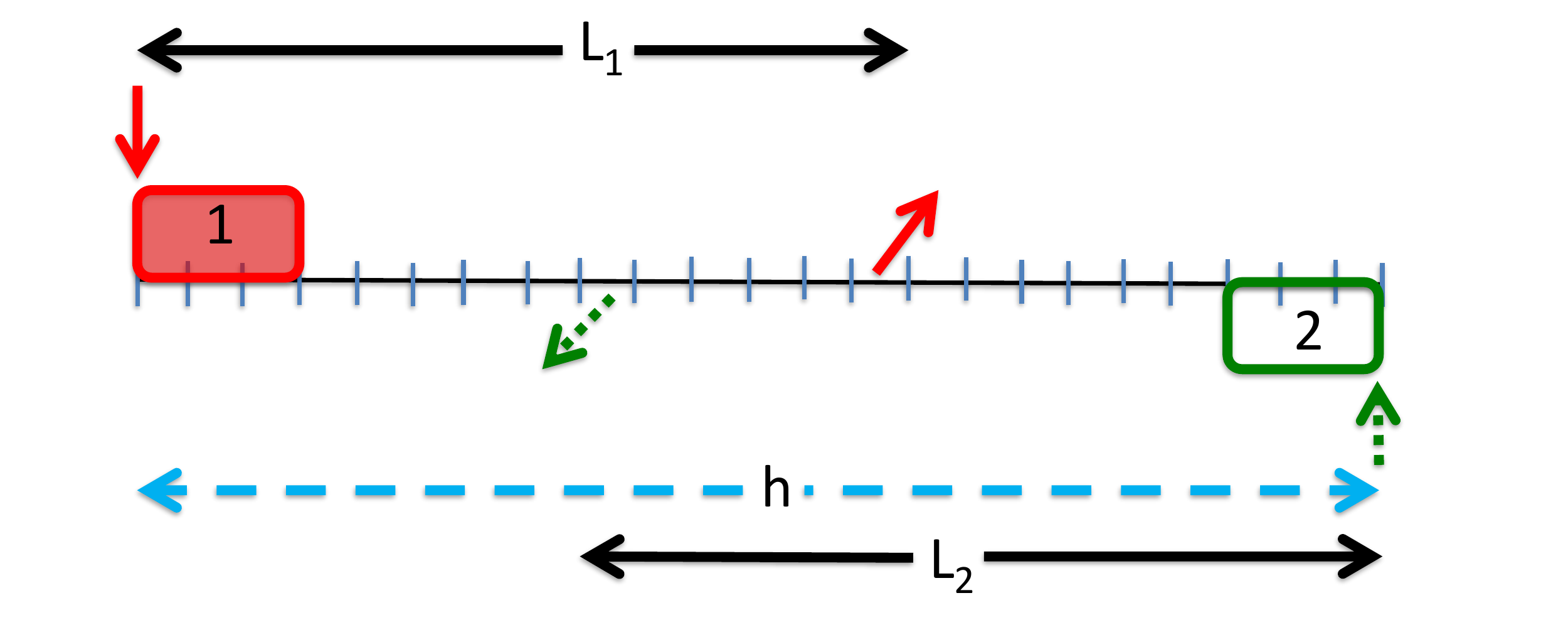}
\caption{(Color online) Tail-to-tail arrangement of the tracks of interfering TASEPs for which 
the master equations governing contra-directional TaI are 
given by eqns.(\ref{eqs:3}) and (\ref{eqs:4}). 
}
 \label{fig-explainEq1}
\end{figure}


In Fig.\ref{fig-switchCO}(a) we plot $J_{1}$ and $J_{2}$ as functions 
of $\alpha_{1}$.  First of all, note that for the selected value of 
$\alpha_2$, flux $J_{2}$ would be fairly high if the rods of species 1 
were not interfering with it. As long as $\alpha_1$ is not too high, 
$J_{2}$ is weakly affected primarily because of the infrequent 
co-directional close encounters (``road blocks'') of the rods of type 2 
with those of type 1. But, as $\alpha_1$ increases, the time gap detected 
at the ON-ramp of species 2 between the departure of a rod of type 1 and 
the arrival of the next rod of the same type becomes shorter. Therefore, 
the ON-ramp of species 2, which is located on the path of the rods of type 
1, remains ``occluded'' for most of the time if $\alpha_1$ is sufficiently 
high. Consequently, a high value of $J_{1}$ strongly suppresses the flux 
$J_{2}$, irrespective of the actual numerical value of $\alpha_{2}$. 

Thus, the fluxes of the two interfering species of rods are strongly 
anti-correlated, and leads to the switch-like behavior. 
Moreover, for a given size ${\ell}$ of the rods, entry of a rod of type 2 at 
its ON-ramp requires successive ${\ell}$ sites at this location must be 
empty. The likelihood of the occurrence of this situation decreases with 
increasing ${\ell}$. Therefore, the suppressive effect of the flux of species 1 
on that of the species 2 is stronger for longer ${\ell}$ (although we have 
verified this fact, no data is being presented here).

\begin{figure}[!ht]
(a) \\
\includegraphics[angle=0,width=0.75\columnwidth]{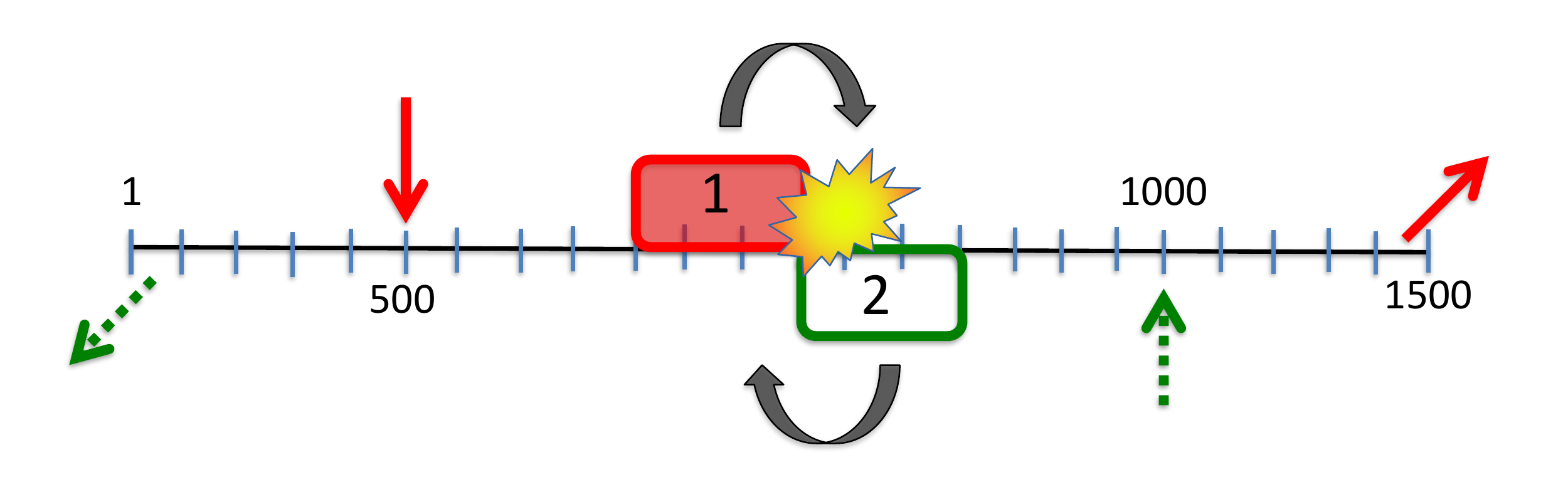}\\
(b) \\
\includegraphics[angle=0,width=0.7\columnwidth]{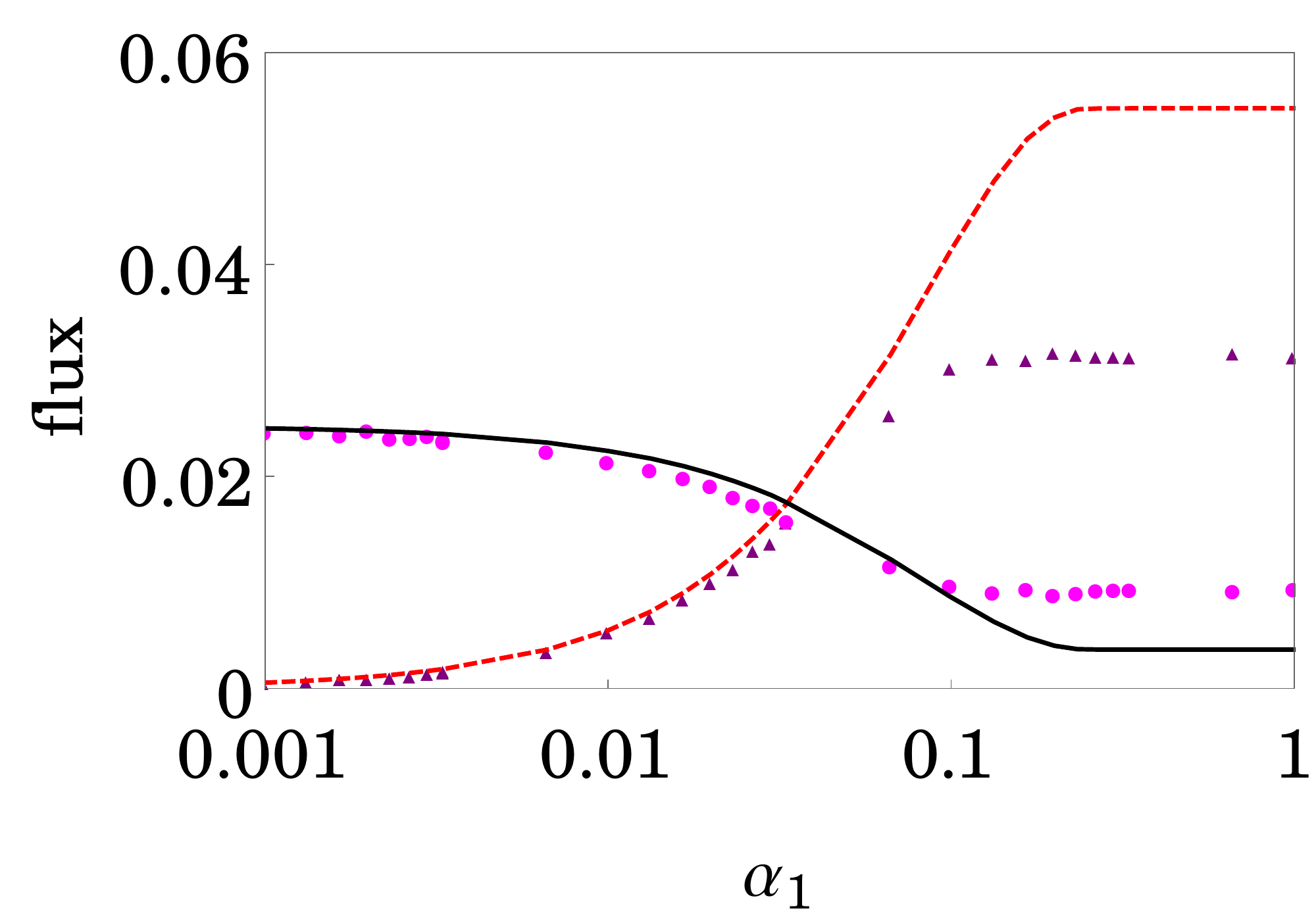}\\
(c)\\
\includegraphics[angle=0,width=0.7\columnwidth]{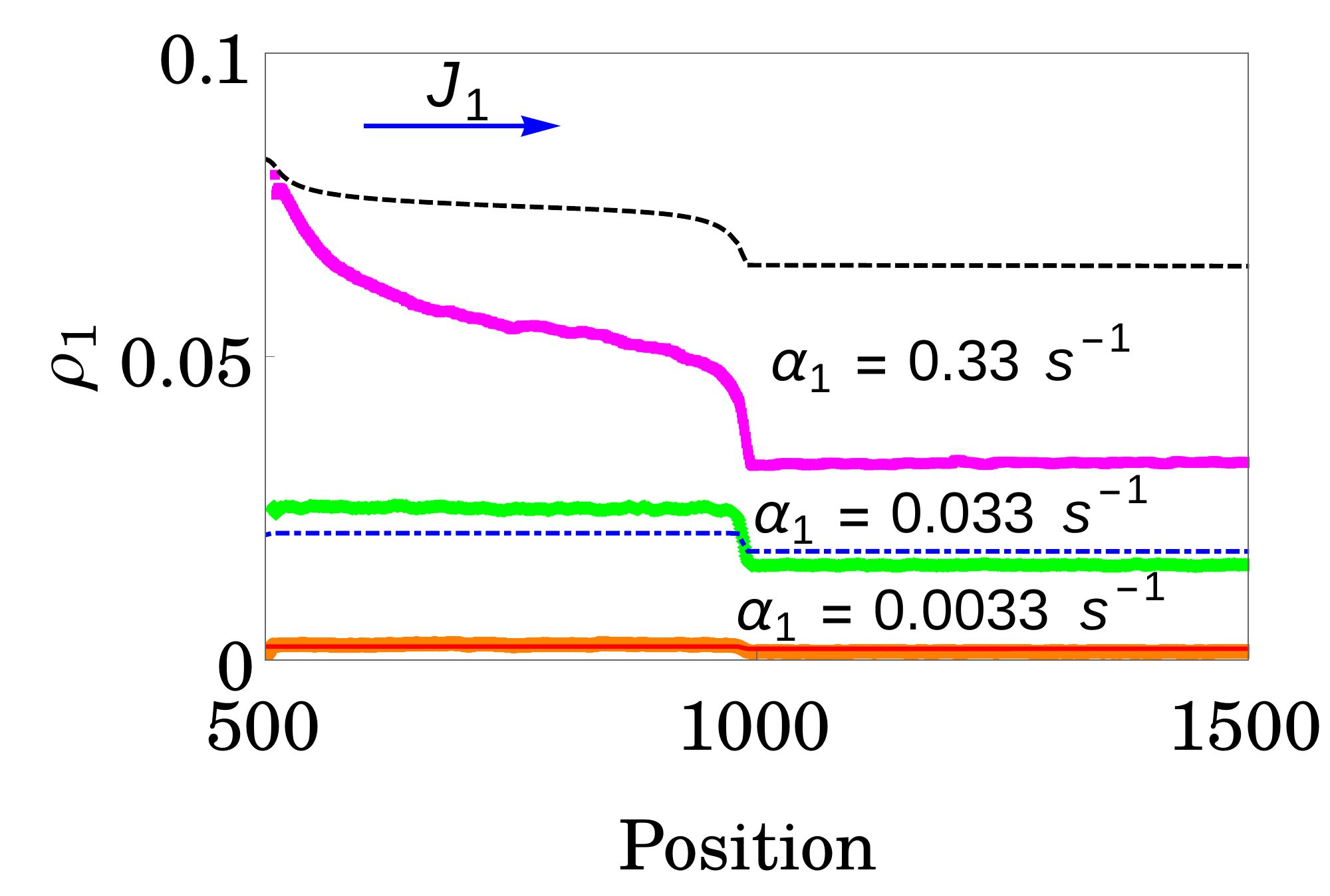}\\
(d)\\
\includegraphics[angle=0,width=0.7\columnwidth]{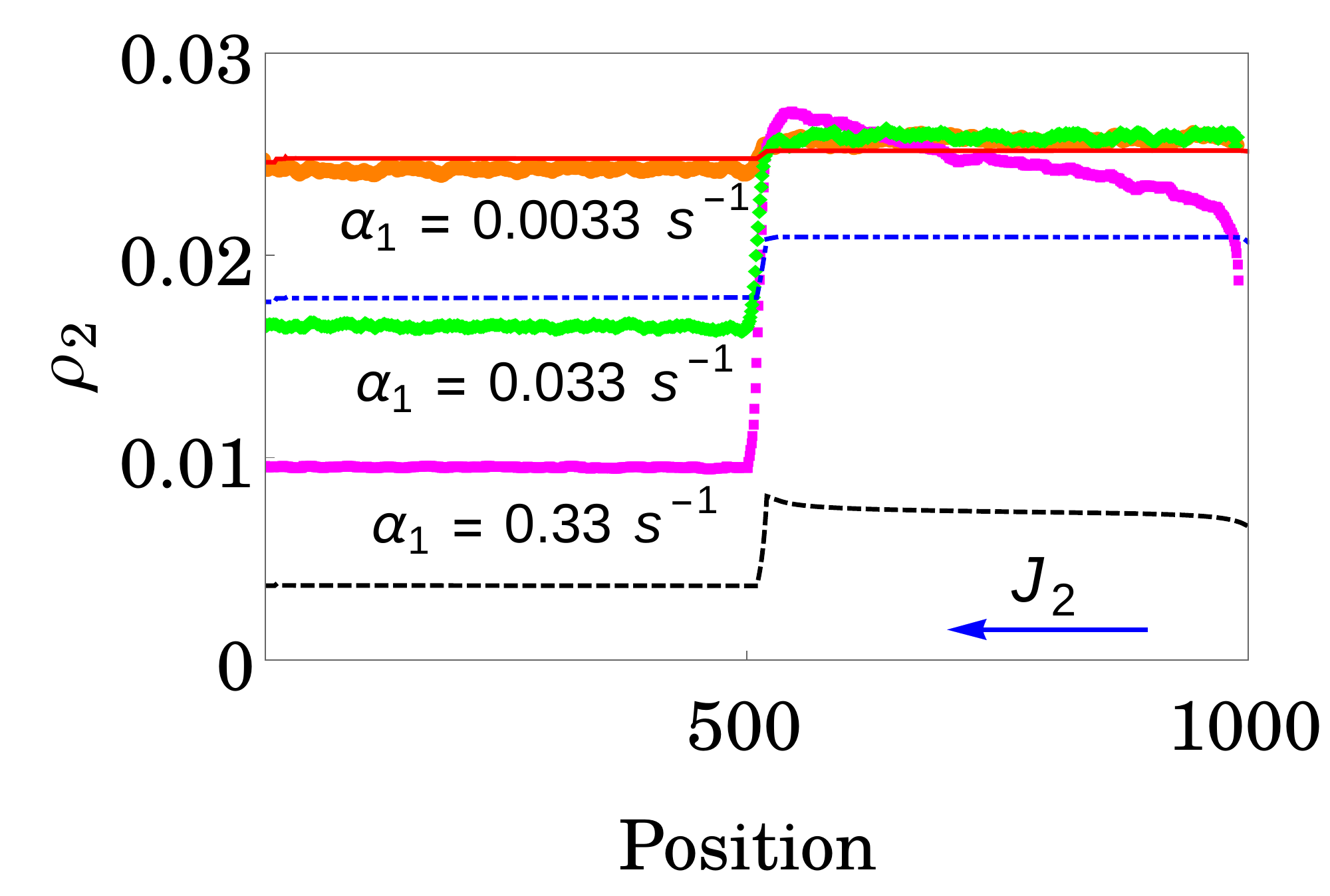}
\caption{(Color online) (a) A schematic representation of contra-directional TaI with passing without detachment on a Head-to-Head arrangement of tracks:  (b) The switch like behaviour of fluxes of rods, plotted as a function of $\alpha_{1}$, for $\beta_{1}=\beta_{2}= 0.33$, $h= 500$, $\alpha_{2} = 0.033$,  $q_{1}=q_{2}= 0.33$ and $\omega_{d1}=\omega_{d2}= 0$. The average density profiles 
$\rho_{1}$ and $\rho_{2}$, for three different values of ${\alpha}_{1}$ are plotted in (c) and (d), 
respectively; the lines and discrete data points correspond to  mean-field theory and computer 
simulations.
}
\label{fig-H2HVir}
\end{figure}	

\begin{figure}[!t]
(a) \\
    \includegraphics[width=0.85\columnwidth]{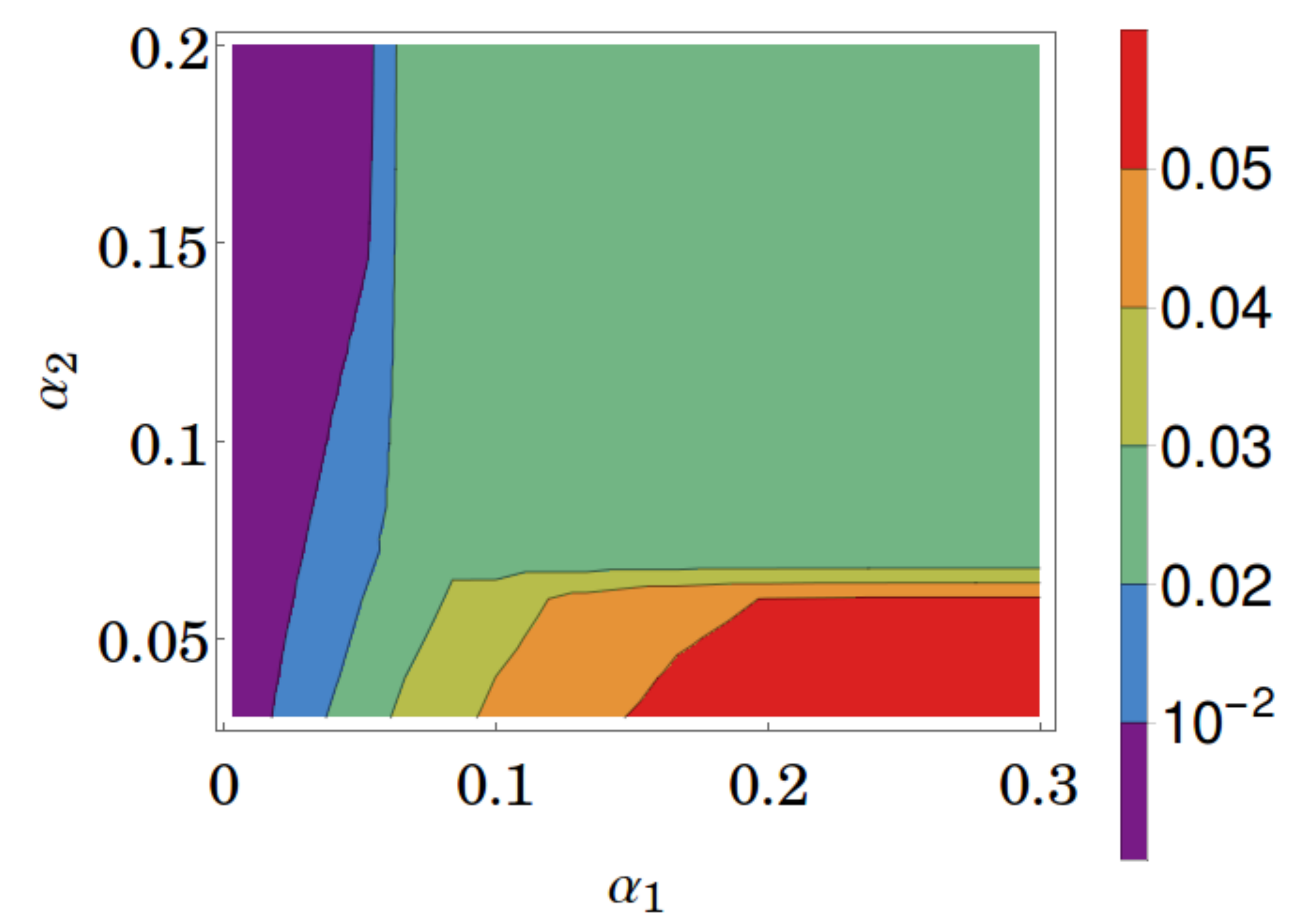}\\
(b)\\
    \includegraphics[width=0.85\columnwidth]{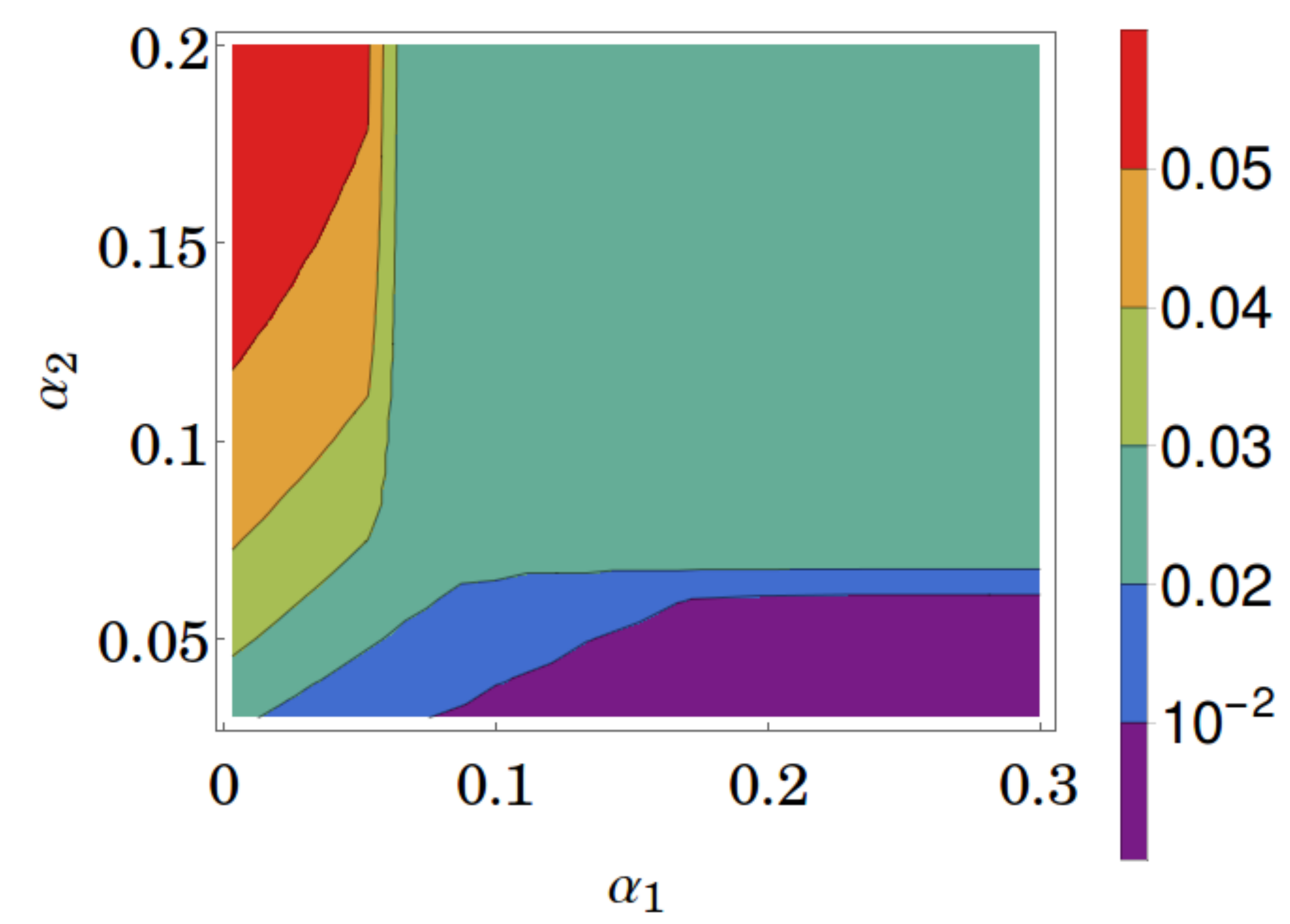}
  \caption{(Color online) Cotradirectional TaI with passing without detachment: Contour plots of (a) $J_{1}$ and (b) $J_{2}$ in the $\alpha_{1}-\alpha_{2}$ plane obtained from mean-field theory for $\beta_{1}=\beta_{2}=\beta = 0.33$. Contours of constant flux is painted by the same color.
}
  \label{fig-ContourContradir}
\end{figure}

\begin{figure}[t]
(a)\\
\includegraphics[angle=0,width=0.8\columnwidth]{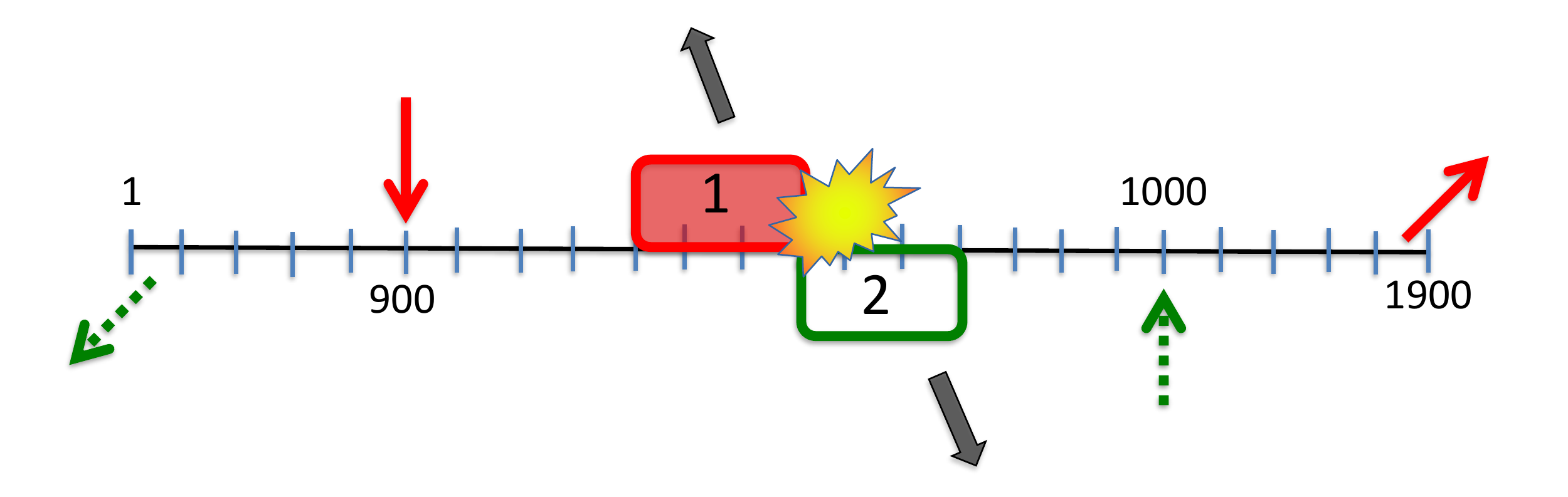}\\
(b)\\
\includegraphics[angle=0,width=0.7\columnwidth]{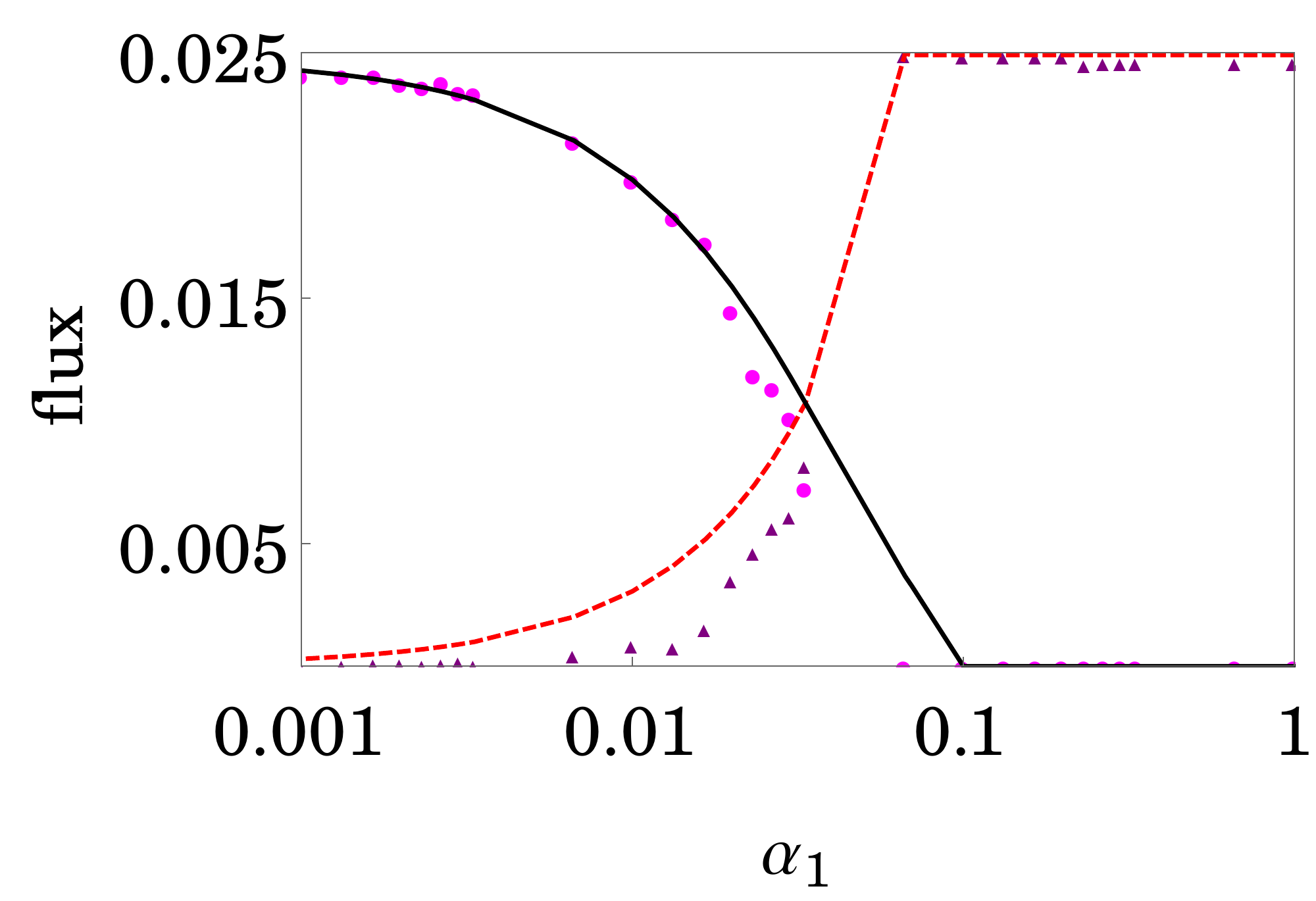}\\
(c)\\
\includegraphics[angle=0,width=0.7\columnwidth]{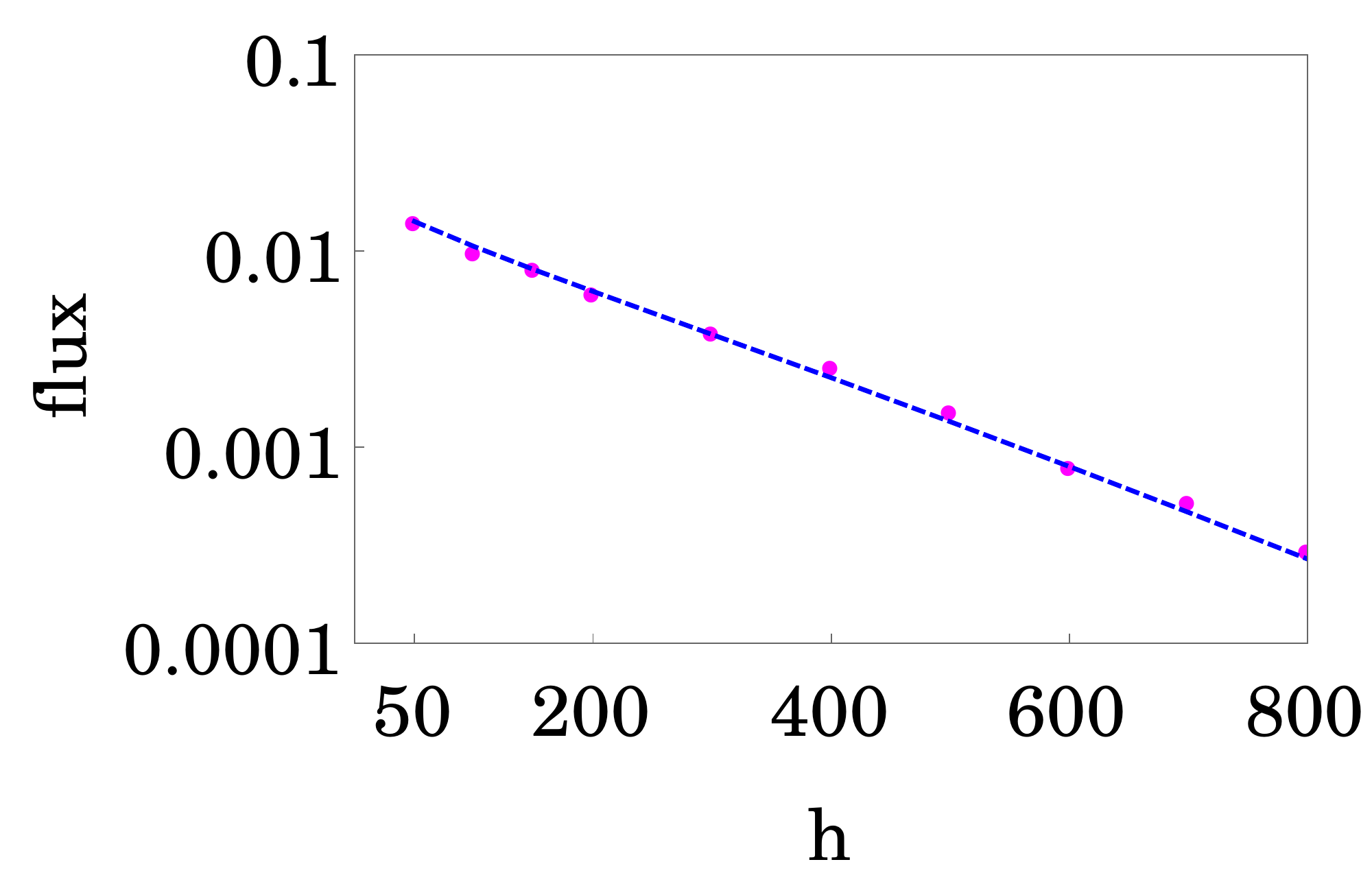}\\
(d)\\
\includegraphics[angle=0,width=0.7\columnwidth]{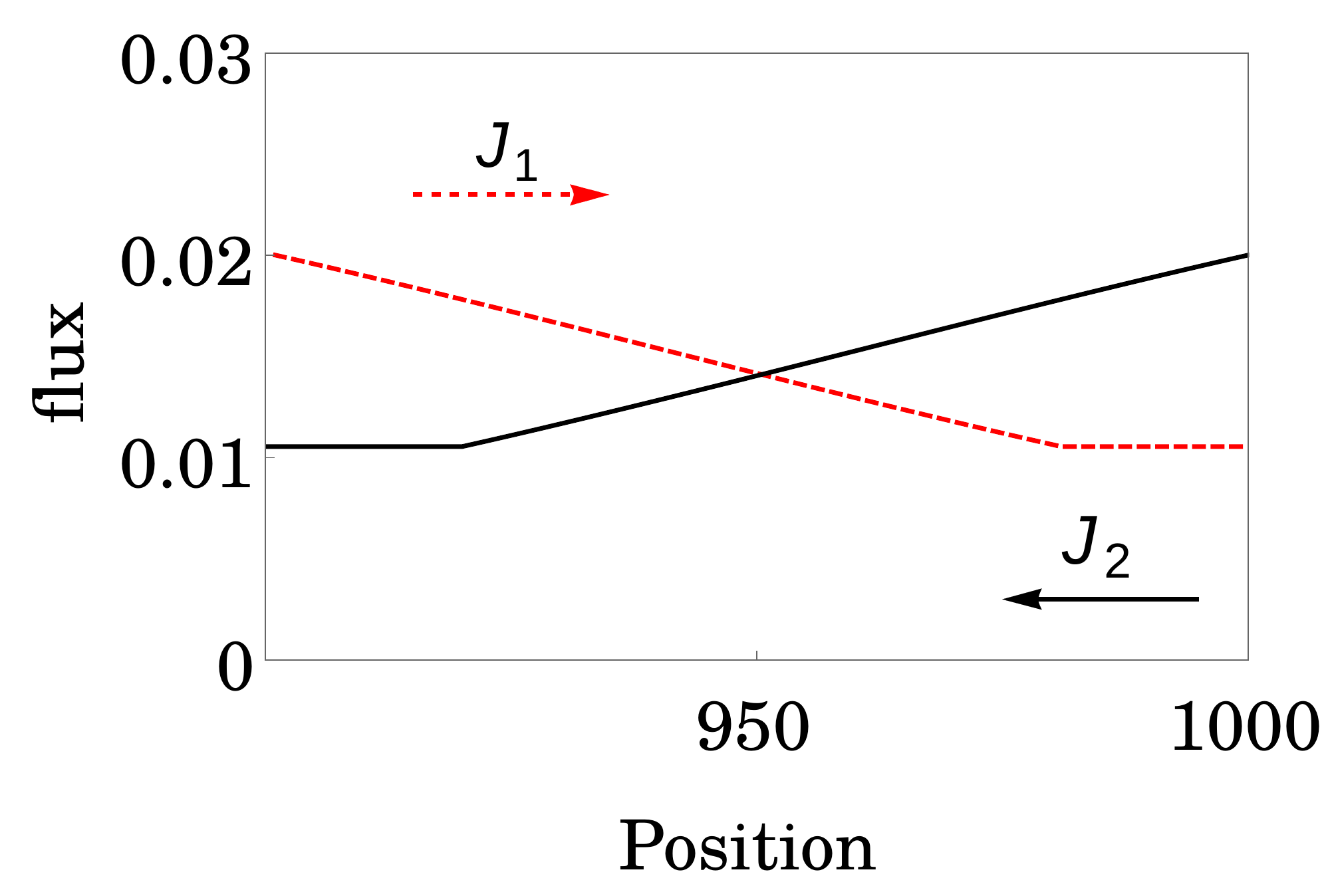}\\
\caption{(Color online) (a) A schematic representation of contra-directional TaI with detachment without passing on a Head-to-Head arrangement  of the tracks. (b) The switch like behaviour of fluxes of rods, plotted as functions of $\alpha_{1}$, for  $h= 100 ~bp$,  $\alpha_{2} = 0.033$, $\beta_{1}=\beta_{2}= 0.033$,  $q_{1}=q_{2}= 0$ and $\omega_{d1}=\omega_{d2}= 0.33$. (c) The semi-log plot of the common current at the point of intersection of the two curves as a function of $h$. (d) The profiles of $J_{1}$ and $J_{2}$ show the decrease of flux in the respective directions of movement of the rods because of detachments suffered as a result of head-on collisions.
}
\label{fig-h2hNV1}
\end{figure}	

The trends of variation of $J_{1}$ and $J_{2}$ with $\alpha_{1}$ are 
consistent with the corresponding variations of the density profiles $\rho_{1}$ 
and $\rho_{2}$ shown in Figs.\ref{fig-switchCO}(b) and (c), respectively.  At 
sufficiently high values of $\alpha_{1}$ the density of the particles of type 2 
becomes vanishingly small throughout the system and the corresponding 
flux is practically zero. At such high values of $\alpha_{1}$ the two-species 
exclusion process is reduced to, effectively, an exclusion process of only the 
type 1 rods. For the specific set of fixed parameters values chosen for this figure the 
system exhibits the maximal current (MC) phase at the high values of $\alpha_{1}$. 
The increasing deviation of the mean-field theoretic prediction from the corresponding 
simulation data in Fig.\ref{fig-switchCO}(a)  is a manifestation of the well known fact 
that mean-field yields a poor approximation for the flux in the MC phase.

More specifically, in the case of a TASEP for single species of rods of length 
${\ell}$ the maximal current (MC) phase occurs \cite{lakatos03,shaw03,shaw04b}  
in the regime $\alpha > \alpha^{*}$, 
$\beta > \beta^{*}$ where $\alpha^{*} = \beta^{*}=\frac{1}{\sqrt{\ell}+1}=0.2403$. 
In this MC phase, i.e., for $(\alpha,\beta)\geq(\alpha^{*},\beta^{*}) $, the 
flux is given by $J_{MC}=\frac{1}{(\sqrt{\ell}+1)^{2}}=0.0577$ and the 
density at the system mid-point, i.e., $\rho_{\frac{N}{2}}=\frac{1}{\sqrt{\ell}(\sqrt{\ell}+1)}=0.076$. 
These numerical values are consistent with the values of $J_{1}$ and $\rho_{1}(L/2)$ 
for the largest values of $\alpha_{1}$ in Figs.\ref{fig-switchCO}(a) and (b), 
respectively. The contours of constant flux $J_{1}$ and those of constant flux $J_{2}$,
plotted in Figs.\ref{fig-ContourCOdir}(a) and (b), respectively, are also consistent with 
this scenario and depict how the two fluxes vary with the variation of $\alpha_{1}$ and 
$\alpha_{2}$ for given $\beta_{1}=\beta_{2}=\beta$.

\section{Contra-directional traffic}
\label{sec-contradir}

In the case of contra-directional TaI, the rods of the two species jump forward 
with different rates $Q_{1}$ and $Q_{2}$, respectively, when there is no 
obstruction in front. However, when two rods of different species face each 
other head-on, two distinct consequences can be envisaged. These two 
distinct scenarios correspond to two different limiting cases of the general 
model of contra-directional TaI that we report in the next two subsections.

In the first plausible scenario, two rods approaching each other  can pass slowly, 
i.e., with a hopping rate that is lower than that in the absence of the obstruction, 
i.e., $q_{1} < Q_{1}$ and $q_{2} < Q_{2}$  (see Fig.\ref{fig-model}(b)). In contrast, 
in the second distinct scenario, up on similar head-on encounter either (or both) 
of the rods get dislodged from the track, at the rates $\omega_{d1}$ and 
$\omega_{d2}$, respectively (see Fig.\ref{fig-model}(c)). Each rod that suffers 
such collision induced detachment from the track fails to reach its designated 
OFF-ramp. In both cases of contra-directional TaI, however, rods of the same 
species are not allowed to pass each other. Biological examples of each of 
these two special limiting cases are mentioned in the preceding section, and the 
corresponding quantitative results are presented in detail, in the next two subsections.

Let $\xi_{1}(\underline{i}|i+\ell)$ be the conditional probability that, 
given that the site $i$ is occupied by (the left edge of ) a rod of species 1, site $i+\ell$ 
is empty, i.e., not covered by any rod. Under mean-field approximation we get 
\begin{eqnarray}
\xi_{1}(\underline{i}|i+\ell)&=&\frac{1 - \sum\limits_{s=1}^{\ell} P_1(i+s)}{1 +P_1(i+\ell) - \sum\limits_{s=1}^{\ell} P_1(i+s)
}
\end{eqnarray}
Similarly, $\xi_{2}(i-\ell|\underline{i})$ is the conditional probability that, 
given that the site $i$ is occupied by (the right edge of ) a rod of species 2 , site $i-\ell$ 
is empty, i.e., not covered by any rod. Under mean-field approximation,
\begin{eqnarray}
\xi_{2}(i-\ell|\underline{i})&=&\frac{1-\sum\limits_{s=1}^{\ell} P_2(i-s)}{1 +P_2(i-\ell) - \sum\limits_{s=1}^{\ell} P_2(i-s)} 
\end{eqnarray}

Let $\xi_{1}(i)$ be the probability that site $i$ on lattice 1 is not 
covered by any rod, irrespective of the state of occupation of any other 
site.  obviously,  $\xi_{1}(i) =  1-\sum_{s=0}^{{\ell}-1}P_{1}(i-s)$.
Note that, if site $i$ is given to be occupied by the left edge of one rod 
of species 1, 
the site $i-1$ on same lattice can be covered by another rod of the 
same species if, and only if, the site $i-{\ell}$ is also occupied by the left edge 
of another rod of species 1.  
Similarly, $\xi_{2}(i)$, the probability that site $i$ on lattice 2 is not 
covered by any rod, irrespective of the state of occupation of any other 
site, is given by  $\xi_{2}(i) = 1-\sum_{s=0}^{{\ell}-1}P_{2}(i+s)$.
Note that, if site $i$ is given to be occupied by the right edge of a rod 
of species 2, 
the site $i+1$ can be covered by another rod of the same species if, and 
only if, the site $i+{\ell}$ is also occupied by the right edge of species 2.

Under MFA, the master equations for the case of tail-to-tail arrangement 
of the tracks ($h>L_1, h>L_2$) (see Fig.\ref{fig-explainEq1}) are written as
\begin{widetext}
\begin{eqnarray}
  \frac{d P_1 \left(1,t\right)}{dt}&=&\alpha_1 ~\xi_{2}(1) ~\xi_{2}(\ell) \Biggl(1-\sum\limits_{s=1}^{\ell}P_1\left(s,t \right) \Biggr) - P_1(1,t)\xi_{1}(\underline{1}|1+{\ell})\left[Q_1\xi_2(1+\ell)+q_1\{1-\xi_2(1+\ell)\}\right]\nonumber\\ &-& P_1(1,t)~\omega_{d1}~P_2(2\ell,t)~,\nonumber \\
  \frac{d P_1\left(i,t\right)}{dt}&=&P_1(i-1,t)\xi_{1}(\underline{i-1}|i-1+{\ell})\left[Q_1\xi_2(i-1+\ell)+q_1\{1-\xi_2(i-1+\ell)\}\right] \nonumber \\
  &-& P_1(i,t)\xi_{1}(\underline{i}|i+{\ell})\left[Q_1\xi_2(i+\ell)+q_1\{1-\xi_2(i+\ell)\}\right]\nonumber\\ &-& P_1(i,t)~\omega_{d1}~P_2(i+2\ell-1,t)~~~{\rm ~for~}, ~(1<i<L_1)~,\nonumber \\
  \frac{d P_1\left(L_1,t\right)}{dt}&=&P_1(L_1-1,t)\xi_{1}(\underline{L_1-1}|L_1-1+{\ell})\left[Q_1\xi_2(L_1-1+\ell)+q_1\{1-\xi_2(L_1-1+\ell)\}\right]\nonumber\\ &-& \beta P_1\left(L_1,t\right)~.\nonumber \\
\label{eqs:3}
\end{eqnarray}

\begin{eqnarray}
  \frac{d P_2 \left(1+h,t\right)}{dt}&=&\alpha_2  ~\xi_{1}(1+h) ~\xi_{1}(1+h-\ell+1) \Biggl(1-\sum\limits_{s=0}^{\ell-1}P_2\left(1+h-s,t \right) \Biggr)\nonumber\\ &-& P_2(1+h,t)\xi_{2}(1+h-\ell|\underline{1+h})\left[Q_2\xi_1(1+h-{\ell})+q_2\{1-\xi_1(1+h-{\ell})\}\right]\nonumber\\ &-& P_2(1+h,t)~\omega_{d2}~P_1(1+h-2\ell+1,t)~,\nonumber \\
   \frac{d P_2\left(i,t\right)}{dt}&=&P_2(i+1,t)\xi_{2}(i+1-\ell|\underline{i+1})\left[Q_2\xi_1(i+1-{\ell})+q_2\{1-\xi_1(i+1-{\ell})\}\right]\nonumber \\ 
  &-& P_2(i,t)\xi_{2}(i-\ell|\underline{i})\left[Q_2\xi_1(i-{\ell})+q_2\{1-\xi_1(i-{\ell})\}\right]\nonumber\\ &-& P_2(i,t)~\omega_{d2}~P_1(i-2\ell+1,t)~~~{\rm ~for~}, ~(2+h-L_2<i<1+h)~, \nonumber \\
  \frac{dP_2\left(2+h-L_2,t\right)}{dt}&=&P_2(3+h-L_2,t)\xi_{2}(3+h-L_2-\ell|\underline{3+h-L_2})\times\nonumber\\&&\left[Q_2\xi_1(3+h-L_2-{\ell})+q_2\{1-\xi_1(3+h-L_2-{\ell})\}\right]\nonumber \\ &-& \beta P_2\left(2+h-L_2,t\right)~. \nonumber \\ 
\label{eqs:4}
\end{eqnarray}
\end{widetext}
Equations for head-to-head arrangement of the two tracks ($h<L_1, h<L_2$) 
are given in the appendix.

\begin{figure}[t]
(a)\\
\includegraphics[angle=0,width=0.8\columnwidth]{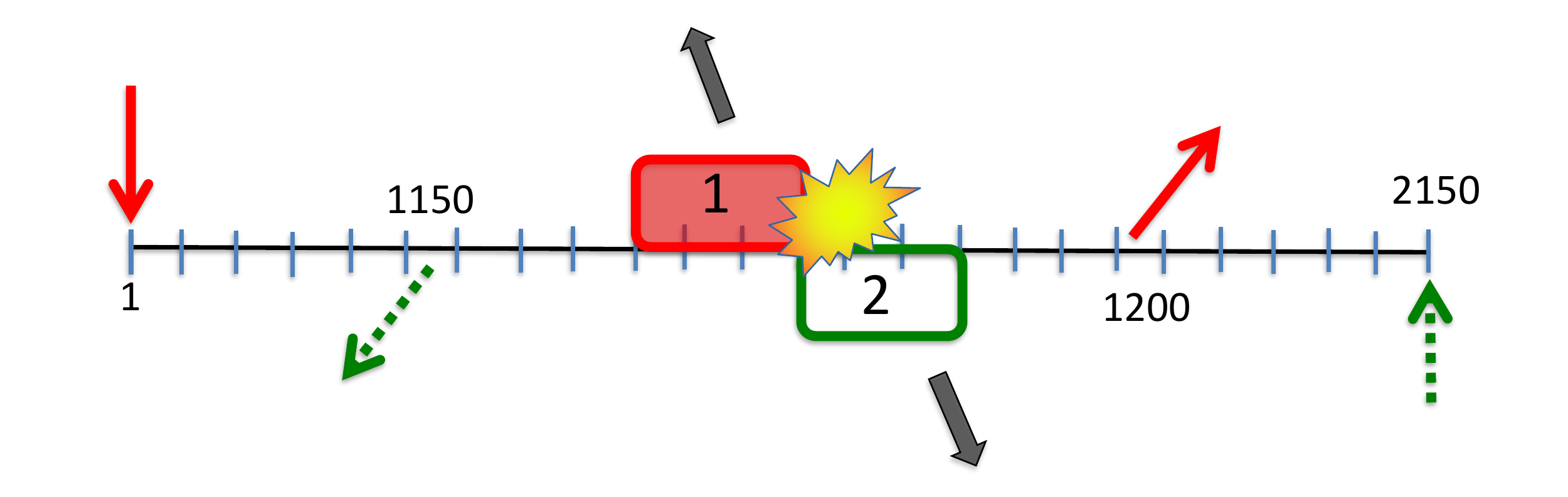} \\
(b)\\
\includegraphics[angle=0,width=0.7\columnwidth]{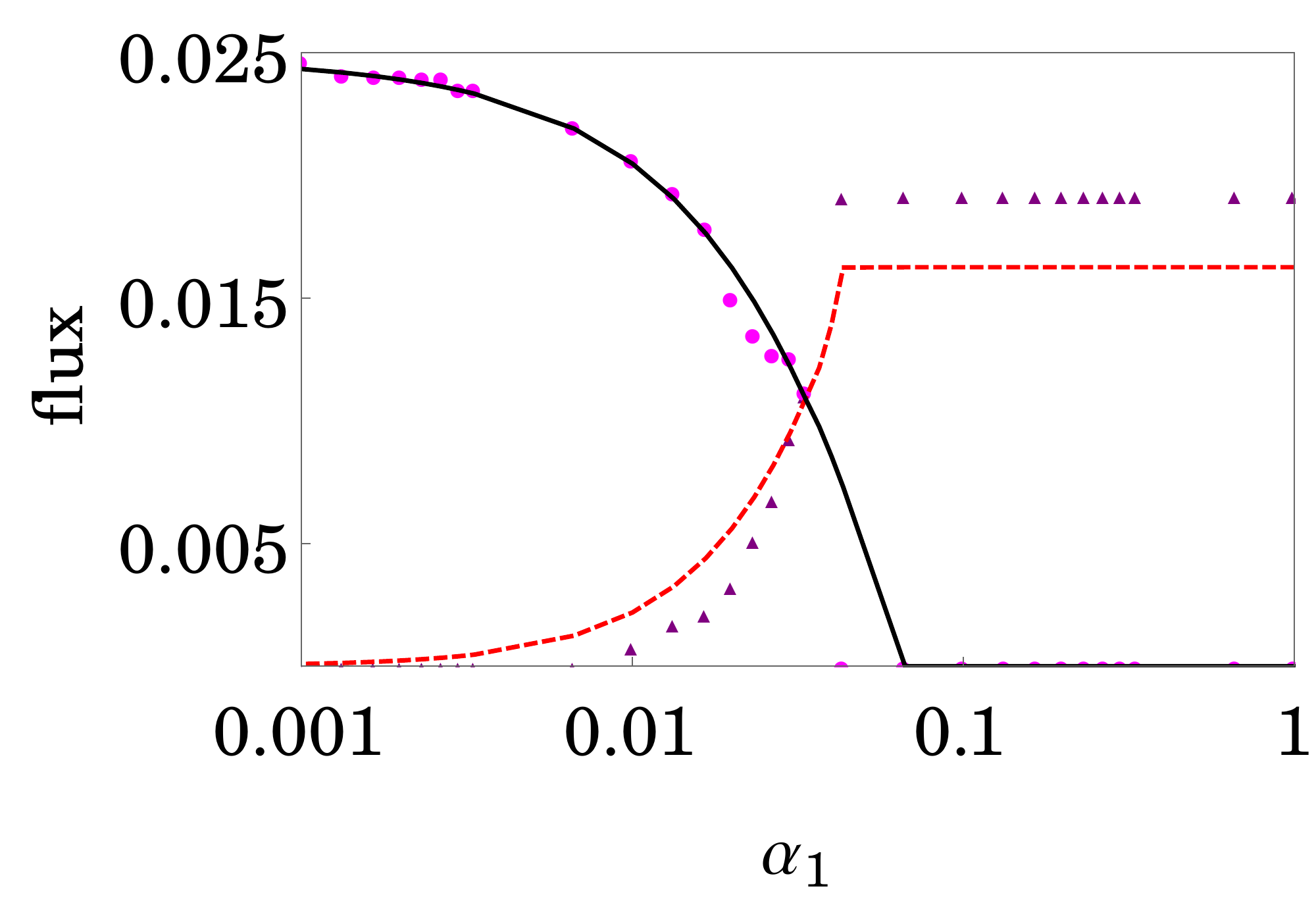}\\
(c) \\
\includegraphics[angle=0,width=0.7\columnwidth]{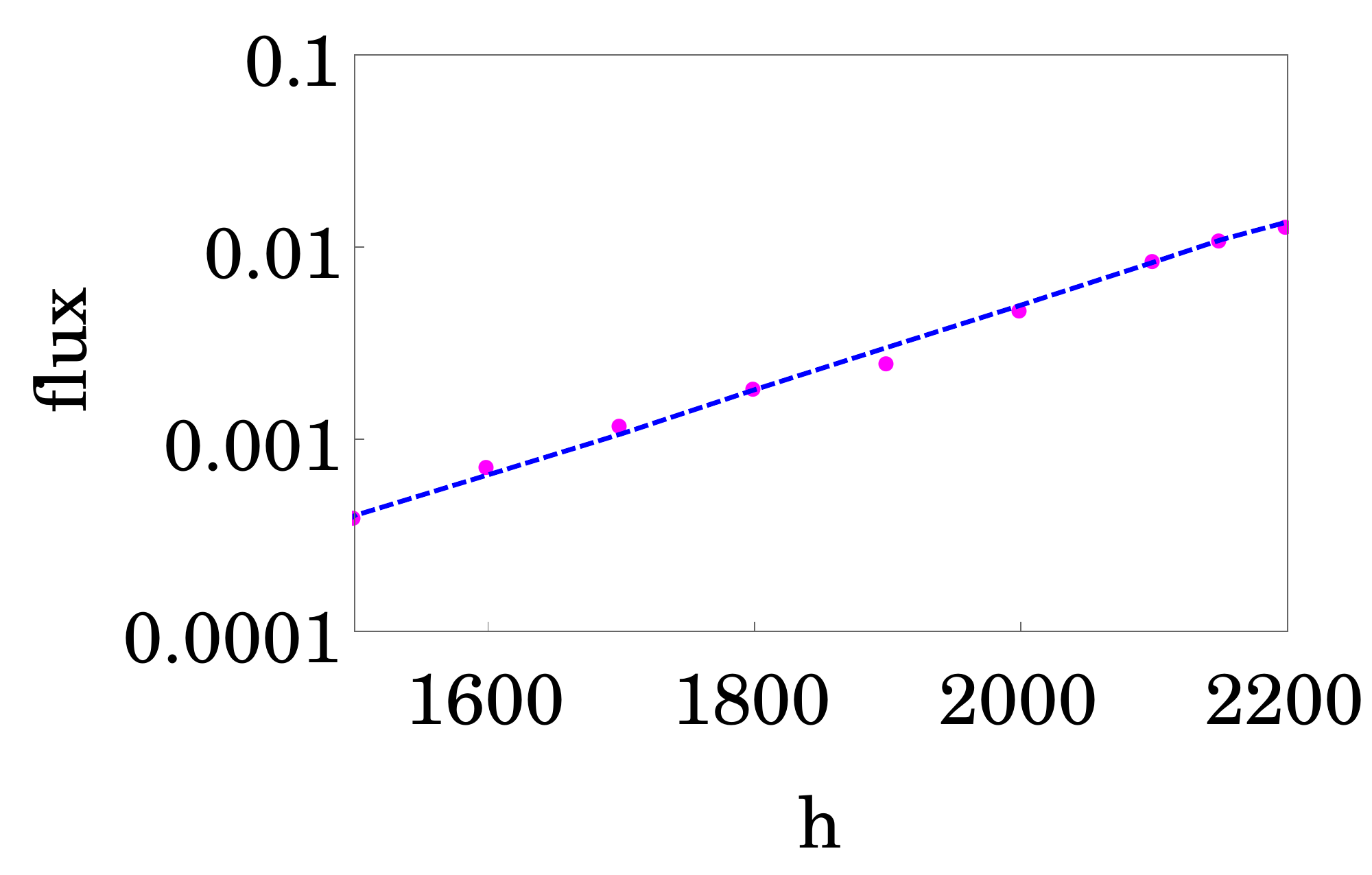}\\
(d) \\
\includegraphics[angle=0,width=0.7\columnwidth]{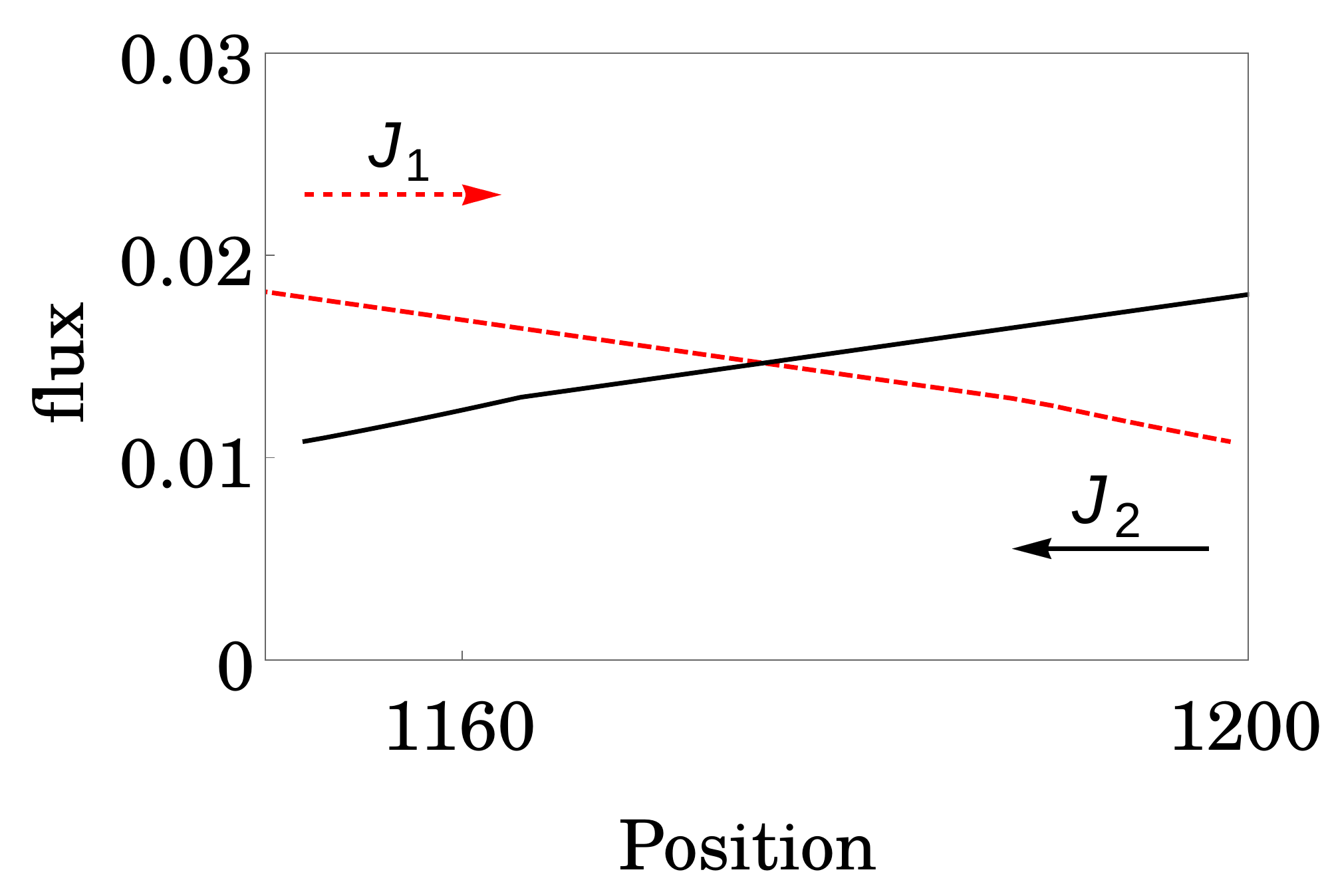}\\
\caption{(Color online) (a) A schematic representation of contra-directional TaI with detachment without passing for Tail-to-tail arrangement of the tracks.
(b) The switch like behaviour of fluxes of rods, plotted as a function of $\alpha_{1}$, for $h= 2150$, 
$\alpha_{2} = 0.033$, $\beta_{1}=\beta_{2}= 0.033$,  $q_{1}=q_{2}= 0$ and $\omega_{d1}=\omega_{d2}= 0.33$. (c) The semi-log plot of the common current at the point of intersection of the two curves as a function of $h$. (d) The profiles of $J_{1}$ and $J_{2}$ show the decrease of flux in the respective directions of movement of the rods because of detachments suffered as a result of head-on collisions.
}
\label{fig-model2}
\end{figure}	

\subsection{Contra-directional TaI with passing without detachment}

In this subsection we present results for the special case 
$\omega_{d1} = 0 = \omega_{d2}$, $q_{1} \neq 0$, $q_{2} \neq 0$. 
This special case is motivated by the experimental observation 
\cite{ma09} that in a head-on collision two  bacteriophage 
(a virus that invades bacteria) RNAPs, approaching each other 
along two different strands of a duplex DNA, can pass. Therefore, 
in this special case of our model, two rods, upon head-on 
encounter, are allowed to pass (see Fig.\ref{fig-H2HVir}(a)), albeit with a 
hopping rate that is lower than that in the absence of the obstruction, i.e., 
$q_{1} < Q_{1}$ and $q_{2} < Q_{2}$. In the context of transcriptional 
interference this slowing down during passing might be caused by the 
mutual hindrance of the RNAPs as well as by the transient structural 
alternation of the macromolecular complex \cite{ma09}.

The mean-field theoretical predictions and the simulation data are plotted 
in Fig.\ref{fig-H2HVir}. The results are qualitatively similar to those plotted 
earlier in Fig.\ref{fig-switchCO} for co-directional TI. However, for the set of 
parameter values used in this figure, there is one feature of the flux and 
density profile that is quantitatively different from those in  Fig.\ref{fig-switchCO}. 
The flux $J_{2}$ in Fig.\ref{fig-H2HVir}(b) saturates to a non-zero value 
with increasing $\alpha_{1}$, instead of vanishing completely, even though 
$J_{1}$ attains the value allowed in the MC phase. The density profile $\rho_{1}$ 
in Fig.\ref{fig-H2HVir}(c), indeed, shows the signature of the MC phase 
corresponding to the largest values of $\alpha_{1}$ while 
$\rho_{2}$ attains a relatively much lower profile (see Fig.\ref{fig-H2HVir}(d)). 
Finally, this scenario is consistent with the contour disgrams in the 
$\alpha_{1}-\alpha_{2}$ plane as depicted in Fig.\ref{fig-ContourContradir}.

\subsection{Contra-directional TaI with detachment without passing}

In this subsection we present results for the special case 
$q_{1}=0=q_{2}$ and $\omega_{d1} \neq 0$, $\omega_{d2} \neq 0$; 
as stated earlier in this paper, these conditions are appropriate 
for modeling transcriptional interference arising from traffic 
of non-bacteriophage RNAPs \cite{shearwin05}.  
Not allowing passing of oppositely moving rods would stall the 
two rods on their respective tracks when these face each other 
head-on. Because of the possibility of detachments of rods up on 
such head-on encounter the flux of the rods would decrease as 
the distance from the ON-ramp increases. The data presented 
in this subsection demonstrate some nontrivial consequences of 
collision-induced premature detachment of rods from their 
respective tracks. In particular, for some specific arrangement 
of the two tracks, the nature of the mutual regulation of the  
TASEPs of two overlapping tracks is very different from that for 
similar track arrangements observed in the preceding subsection.

In case of head-to-head arrangement of the two tracks (see Fig.\ref{fig-h2hNV1}(a)), 
the results presented in Fig.\ref{fig-h2hNV1}(b) demonstrate the switch-like 
regulation of one TASEP by the other.  
Since $\alpha_{2}=0.033$, the two curves for $J_{1}$ and $J_{2}$ 
intersect at $\alpha_{1}=0.033$ for all the values of $h$ for 
which switch-like behavior was observed. However, the actual magnitudes 
of the two fluxes at the point of intersection decreases with increasing 
$h$ (see Fig.\ref{fig-h2hNV1}(c)); this trend of variation is a consequence 
of the gradual attenuation of flux caused by the increasing number of 
rod detachment events (see Fig.\ref{fig-h2hNV1}(d)). 
Since in all those cases  $\beta$ is small, the 
density of the respective rods is expected to be high. Therefore, each 
rod would be involved in frequent head-on encounters each of which is 
a potential cause for its detachment from its track. The longer the 
travel, the higher is the attenuation of flux caused by such premature 
detachments of the rods from their respective tracks resulting in lower 
overall rate of flux measured at the OFF-ramp.

Switch-like regulation of the interfering TASEPs is observed also 
in the case of tail-to-tail arrangement of the tracks (see 
Fig.\ref{fig-model2}). With the increase of $\alpha_{1}$, larger 
number of rods begin their journey from the on-ramp on track 1. 
Moreover, for fixed $L_1$ and $L_2$, the larger the magnitude of  
$h$ the shorter is the overlap between the two tracks. Therefore, 
with increasing $h$ fewer head-on collisions are expected. 
However, as long as the overlap is non-zero and $\beta=0.033$, 
$\alpha_{2}=0.033$, the head-on collisions are sufficiently 
frequent in the overlap region for all $\alpha_{1} > 0.033$ 
to reduce $J_{2}$ to vanishingly low level (see fig.\ref{fig-model2}(b)). 
Since $\alpha_{1}$ is larger than $\alpha_{2}$ in this regime, 
more and more rods of species 1 successfully reach their 
OFF-ramp without suffering collision-induced premature detachment. 
That is why, as seen in Fig.\ref{fig-model2}(b), $J_{1}$ increases 
with increasing $\alpha_{1}$ in this regime. Finally, at sufficiently 
large values of $\alpha_{1}$, $J_{1}$ attains its saturation value 
for the given set of parameters. This interpretation of the 
results of Fig.\ref{fig-model2}(b) is reinforced by the data 
presented in Figs.\ref{fig-model2}(c) and (d). 

We conclude that when passing is not allowed, the switch-like regulation 
of the two interfering TASEPs is caused primarily by of the premature 
detachments of the rods from the tracks upon {\it head-on encounters}. This 
mechanism is very different from the mechanisms of switching off of TASEP 
that was observed in the preceding subsection where the cause was {\it occlusion}.   

The agreement between the mean-field theory and simulation data in 
Fig.\ref{fig-model2}(b) (tail-to-tail) is not as good as that in Fig.\ref{fig-h2hNV1}(b).
We record here one observation that may be the root cause of this difference. 
In Fig. 8b, which corresponds to Head-to-Head arrangement of the two tracks, 
density of the type 2 rods is very low at high value of $\alpha_{1}$. In contrast, 
in case of Tail-to-Tail arrangement, densities of both types of rods can be high 
even at high values of $\alpha_{1}$, because the two on-ramps are out of the 
region of overlap of the two tracks.

\section{Summary and conclusion}
\label{sec-conclusion}

In this paper we have developed a class of biologically motivated exclusion models 
of two distinguishable species of hard rods, with their respective distinct points of 
entry and exit (ON- and OFF-ramps in the terminology of traffic models). 
The main focus of our theoretical investigation is on the influence of the flux of 
one species of rods on that of the other. We have modelled both co-directional and 
contra-directional traffic of the two species of rods. Moreover, In the case of 
contra-directional TaI, we have considered both head-to-head and tail-to-tail 
arrangement of the tracks of the two TASEPs. Furthermore, 
in the case of contra-directional traffic of the TASEPs the results for 
two special cases have been presented separately; in one case two rods, 
upon head-on collision, can pass each other without detachment from the 
track whereas in the other the rods cannot pass but can detach from the track.

We have analyzed the kinetics of the models (a) theoretically under mean-field 
approximation (MFA) and (b) computationally by computer (Monte Carlo) simulation. 
Except for  parameter regimes where significant correlations exist, that are neglected 
by MFA, the predictions of the approximate mean-field theory are in excellent agreement 
with the simulation data. 
Under wide varieties of conditions the system exhibits a bistable switch-like behavior: 
switching ON a high rate of entry of one species of rods at its ON-ramp switches 
OFF the flux of the other species of rods measured at the OFF-ramp of the latter. 
Besides, for some set of values of the parameters,  the suppressive effect of even 
the highest level of flow of one  of rods is not strong enough to completely switch OFF 
the flux of the other species although the latter suffers significant reduction of its flux.

``Occlusion'' in TI is captured by the terms proportional to $\alpha_{1}$ and $\alpha_{2}$ 
in the master equations for our models. The terms involving the conditional probabilities 
$\xi$ capture the effects of ``road blocks''. The terms proportional to $\omega_{d1}$ 
and those proportional to $\omega_{d2}$ account for the effects of ``collision'' resulting 
in premature detachment. In our simplified models the so-called ``sitting duck'' mode of 
TI \cite{shearwin05} (not shown in Fig.\ref{fig-TImodes}) cannot be distinguished from 
collision. If one of the rods detaching from its track upon head-on collision happens to be 
located at its ON-ramp at the instant of collision, it may be interpreted as a physical 
realization of sitting duck mode of TI.  

In case co-directional interference the dominant cause for the switching OFF of the flux 
of one species of rods  is the {\it occlusion} of its ON-ramp by the rods of other species. 
In contrast, in case of the tail-to-tail arrangement of the tracks for contra-directional 
interference none of the ON-ramps are accessible to the oncoming traffic of the other 
species of rods; in this case detachment of rods upon head-on collision is the dominant 
cause of switching OFF the lower flux by the higher flux of the other species of rods. 
Both the occlusion and collision-induced detachments can cause drastic reduction in 
the flux of one species of rods in the case of contra-directional interference if the 
tracks are arranged in a head-to-head fashion; which of these two plays the dominant 
role is decided by the magnitudes of the set of rate constants.

From the perspective of regulation and control, each of the  two species of rods 
regulates the flux of the other in a manner that is very similar to regulation of 
the level of transcription in the smallest ``self-regulatory'' circuit formed by 
just two overlapping genes \cite{pelechano13,georg11,lapidot06}. 
From the detailed exploration of the three scenarios encapsulated by 
Fig.\ref{fig-model} we have also discovered more than one possible 
mechanisms of the switch-like regulation of the fluxes.

\begin{appendix}

\section{Master equations for contra-directional flow in TaI: Head-to-head arrangement of the tracks}

\begin{figure}[h]
  \includegraphics[angle=0,width=0.85\columnwidth]{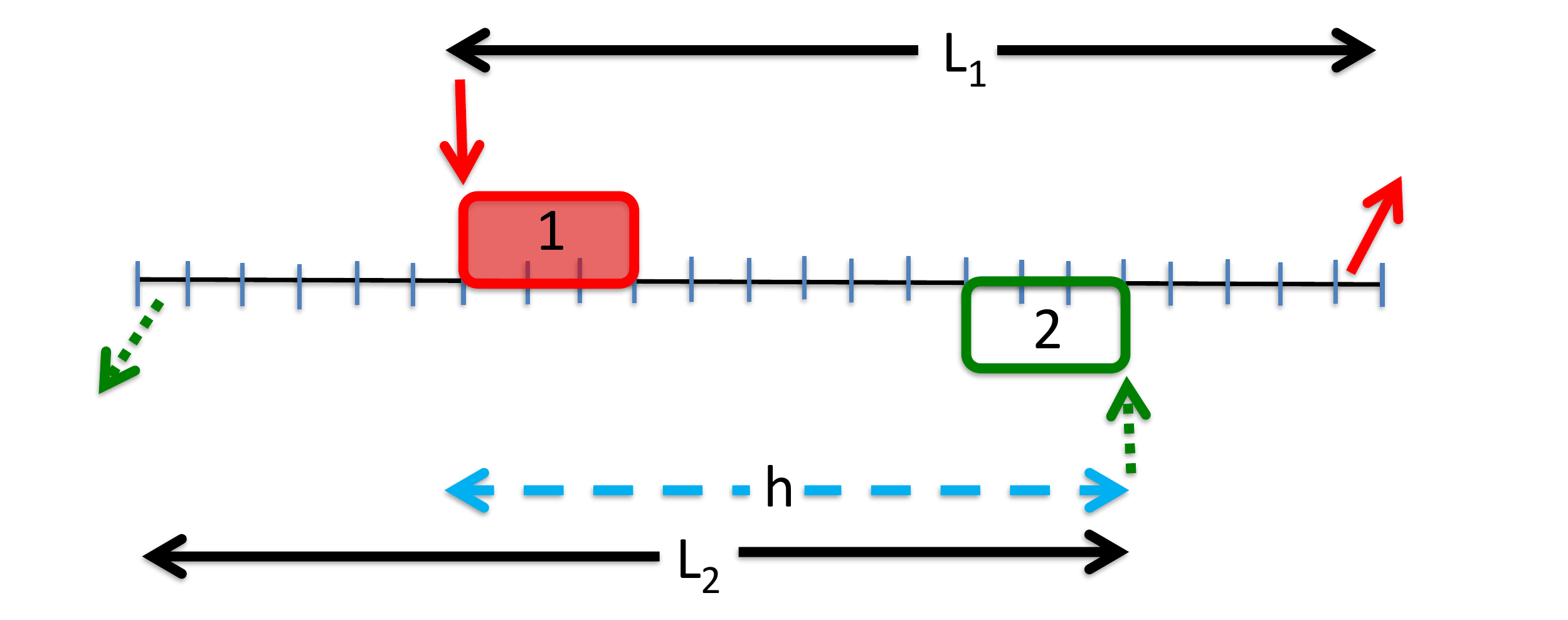}
  \caption{(Color online) Head-to-head arrangement of the tracks of interfering TASEPs for which the master 
equations governing contra-directional rod traffic are given by the  equations (\ref{eqs:5}) and 
(\ref{eqs:6}).
}
  \label{fig-explainEq2}
\end{figure}


Equations for head-to-head arrangment of the tracks ($h<L_1, h<L_2$)(see fig.\ref{fig-explainEq2}) are written as
\begin{widetext}
\begin{eqnarray}
  \frac{d P_1 \left(L_2-h,t\right)}{dt}&=&\alpha_1 ~\xi_{2}(L_2-h) ~\xi_{2}(L_2-h+\ell-1) \Biggl(1-\sum\limits_{s=0}^{\ell-1}P_1\left(L_2-h+s,t \right) \Biggr) \nonumber \\ &-& P_1(L_2-h,t)\xi_{1}(\underline{L_2-h}|L_2-h+{\ell})\left[Q_1\xi_2(L_2-h+\ell)+q_1\{1-\xi_2(L_2-h+\ell)\}\right]\nonumber\\ &-& P_1(L_2-h,t)~\omega_{d1}~P_2(L_2-h+2\ell-1,t)~,\nonumber \\
  \frac{d P_1\left(i,t\right)}{dt}&=&P_1(i-1,t)\xi_{1}(\underline{i-1}|i-1+{\ell})\left[Q_1\xi_2(i-1+\ell)+q_1\{1-\xi_2(i-1+\ell)\}\right] \nonumber \\
  &-& P_1(i,t)\xi_{1}(\underline{i}|i+{\ell})\left[Q_1\xi_2(i+\ell)+q_1\{1-\xi_2(i+\ell)\}\right]\nonumber\\ &-& P_1(i,t)~\omega_{d1}~P_2(i+2\ell-1,t)~~~{\rm ~for~}, ~(L_2-h<i<L_1+L_2-h-1)~,\nonumber \\
  \frac{d P_1\left(L_1+L_2-h-1,t\right)}{dt}&=&P_1(L_1+L_2-h-2,t)\xi_{1}(\underline{L_1+L_2-h-2}|L_1+L_2-h-2+{\ell})\times\nonumber\\&&\left[Q_1\xi_2(L_1+L_2-h-2+\ell)+q_1\{1-\xi_2(L_1+L_2-h-2+\ell)\}\right]\nonumber\\ &-& \beta P_1\left(L_1+L_2-h-1,t\right)~.\nonumber \\
\label{eqs:5}
\end{eqnarray}

\begin{eqnarray}
  \frac{d P_2 \left(L_2,t\right)}{dt}&=&\alpha_2  ~\xi_{1}(L_2) ~\xi_{1}(L_2-\ell+1) \Biggl(1-\sum\limits_{s=0}^{\ell-1}P_2\left(L_2-s,t \right) \Biggr)\nonumber\\ &-& P_2(L_2,t)\xi_{2}(L_2-\ell|\underline{L_2})\left[Q_2\xi_1(L_2-{\ell})+q_2\{1-\xi_1(L_2-{\ell})\}\right]\nonumber\\ &-& P_2(L_2,t)~\omega_{d2}~P_1(L_2-2\ell+1,t)~,\nonumber \\
   \frac{d P_2\left(i,t\right)}{dt}&=&P_2(i+1,t)\xi_{2}(i+1-\ell|\underline{i+1})\left[Q_2\xi_1(i+1-{\ell})+q_2\{1-\xi_1(i+1-{\ell})\}\right]\nonumber \\ 
  &-& P_2(i,t)\xi_{2}(i-\ell|\underline{i})\left[Q_2\xi_1(i-{\ell})+q_2\{1-\xi_1(i-{\ell})\}\right]\nonumber\\ &-& P_2(i,t)~\omega_{d2}~P_1(i-2\ell+1,t)~~~{\rm ~for~}, ~(1<i<L_2)~, \nonumber \\
  \frac{dP_2\left(1,t\right)}{dt}&=&P_2(2,t)\xi_{2}(2-\ell|\underline{2})\left[Q_2\xi_1(2-{\ell})+q_2\{1-\xi_1(2-{\ell})\}\right]\nonumber \\ &-& \beta P_2\left(1,t\right)~. \nonumber \\ 
\label{eqs:6}
\end{eqnarray}
\end{widetext}

\end{appendix}

\section*{Acknowledgements} 
This work is supported by ``Prof. S. Sampath Chair'' Professorship (DC), 
a J.C. Bose National Fellowship (DC) and DST Inspire Faculty Fellowship (TB). 
We also thank Mustansir Barma for useful discussions.


\end{document}